\documentclass[fleqn,pra]{revtex4}

\usepackage{amsmath,graphicx}
\usepackage{verbatim}

\begin{document}

\title{Hyperfine structure in the H$_2^+$ and HD$^+$ molecular ions at $m\alpha^6$ order}

\author{Vladimir I. Korobov$^1$, Jean-Philippe Karr$^{2,3}$, Mohammad Haidar$^2$, and Zhen-Xiang Zhong$^4$}
\affiliation{$^1$Bogoliubov Laboratory of Theoretical Physics, Joint Institute for Nuclear Research, Dubna 141980, Russia}
\affiliation{$^2$Laboratoire Kastler Brossel, Sorbonne Universit\'e, CNRS, ENS-PSL Research University, Coll\`ege de France, 4 place Jussieu, F-75005 Paris, France}
\affiliation{$^3$Universit\'e d'Evry-Val d'Essonne, Universit\'e Paris-Saclay, Boulevard Fran\c cois Mitterrand, F-91000 Evry, France}
\affiliation{$^4$Division of Theoretical and Interdisciplinary Research, State Key Laboratory of Magnetic Resonance and Atomic and Molecular Physics, Wuhan Institute of Physics and Mathematics, Chinese Academy of Sciences, Wuhan 430071, China}

\begin{abstract}
A complete effective Hamiltonian for relativistic corrections at orders $m\alpha^6$ and $m\alpha^6(m/M)$ in a one-electron molecular system is derived from the NRQED Lagrangian. It includes spin-independent corrections to the energy levels and spin-spin scalar interactions contributing to the hyperfine splitting, both of which had been studied previously. In addition, corrections to electron spin-orbit and spin-spin tensor interactions are newly obtained. This allows improving the hyperfine structure theory in the hydrogen molecular ions. Improved values of the spin-orbit hyperfine coefficient are calculated for a few transitions of current experimental interest.
\end{abstract}

\maketitle

\section{Introduction}

High-resolution spectroscopy of the hydrogen molecular ions H$_2^+$ and HD$^+$ may contribute significantly to the determination of fundamental constants such as the proton-electron mass ratio $m_p/m_e$~\cite{Karr16}. A pure rotational transition in HD$^+$ has recently been measured with a relative uncertainty of $1.3 \times 10^{-11}$~\cite{Alighanbari20}. The experimental accuracy of ro-vibrational transition frequencies is expected to reach a few parts per trillion in the near future using spectroscopy in the Lamb-Dicke regime~\cite{Alighanbari20,Tran13,Patra19} or in a Doppler-free geometry~\cite{Karr08,Louvradoux19}. While information on fundamental constants is obtained from comparison of spin-averaged transition frequencies with theoretical predictions, the hyperfine splitting of ro-vibrational lines also allows for precise tests of theory.

So far, the hyperfine structure of H$_2^+$ and HD$^+$ has been calculated within the Breit-Pauli approximation~\cite{Bakalov06,Korobov06}, taking into account the anomalous magnetic moment of the electron. All terms at orders $m\alpha^{4}$ and $m\alpha^5$ are included, so that the theoretical accuracy of the hyperfine coefficients is of order $\alpha^2 \sim 5 \times 10^{-5}$. Higher-order corrections to the largest coefficients, i.e. the spin-spin Fermi contact interaction, were later calculated in~\cite{Korobov09,Korobov16}, which allowed to get excellent agreement with available RF spectroscopy data in H$_2^+$~\cite{Jefferts69} at the level of $\sim 1$~ppm. The following step to improve the hyperfine structure theory is to evaluate higher-order corrections to the next largest coefficients, i.e. the electron spin-orbit and spin-spin tensor interaction, starting with relativistic corrections at the $m\alpha^6$ order.

With this aim, we derive in the present work the complete effective Hamiltonian for the hydrogen molecular ions at the $m\alpha^6$ and $m\alpha^6(m/M)$ orders, following the NRQED approach~\cite{Caswell86,Kinoshita96,Hill13,Haidar20}. Then, we use it to calculate numerically the corrections to the electron spin-orbit interaction for a few transitions studied in ongoing experiments. The paper is organized as follows: in Secs.~\ref{sec-Lagrangian} and~\ref{vertices}, we recall the expression of the NRQED Lagrangian and associated interaction vertices. We then systematically derive the effective potentials, which are organized in three categories: tree-level interactions involving the exchange of a Coulomb or transverse photon~(Sec.~\ref{tree}), terms due to retardation in the transverse photon exchange~(Sec.~\ref{sec-retard}), and finally those coming from a seagull diagram with simultaneous exchange of two photons~(Sec.~\ref{seagull}). In Sec.~\ref{heff6}, we collect our results to write the total effective Hamiltonian, separating the different types of interactions: spin-independent, electronic spin-orbit, spin-spin scalar and tensor interactions. Finally, in Sec.~\ref{results} we present numerical calculations of the spin-orbit interaction coefficient.

\section{NRQED Lagrangian } \label{sec-Lagrangian}

Natural (Lorentz-Heaviside) units ($\hbar=c=1$) are used throughout. We assume that $e$ is the electron's charge and thus is negative, the elementary charge is then denoted by $|e|$.

We use the Coulomb gauge for photons, and electrons are described by two-component Pauli spinors. We take the NRQED Lagrangian for the electron in the gauge-invariant form~\cite{Caswell86,Kinoshita96,Hill13,Haidar20}, including all the terms involved in bound-state energy corrections up to the $m\alpha^6$ order:
\begin{equation}\label{Lagrangian}
\begin{array}{@{}l}\displaystyle
L_{\rm main} =
\psi_e^*\left(i\partial_t-eA_0+\frac{\mathbf{D}^2}{2m}
            +\frac{\mathbf{D}^4}{8m^3}+\frac{\mathbf{D}^6}{16m^5}+\dots\!\right)\psi_e
\\[3mm]\hspace{12mm}\displaystyle
   +\psi_e^*\left(c_F\frac{e}{2m}\boldsymbol{\sigma}\!\cdot\!\mathbf{B}
                   +c_D\frac{e}{8m^2}\Bigl(\mathbf{D}\!\cdot\!\mathbf{E}-\mathbf{E}\!\cdot\!\mathbf{D}\Bigr)
                   +c_S\frac{ie}{8m^2}\,
                   \boldsymbol{\sigma}\!\cdot\!
                   \Bigl(
                      \mathbf{D}\!\times\!\mathbf{E}
                      -\mathbf{E}\!\times\!\mathbf{D}
                   \Bigr)
              \right.
\\[3mm]\hspace{12mm}\displaystyle
   +c_{W_1}\frac{e}{8m^3}\Bigl\{\mathbf{D}^2,\boldsymbol{\sigma}\mathbf{B}\Bigr\}
   +\frac{3ie}{16m^4}
      \Bigl\{
         \mathbf{D}^2,
         \boldsymbol{\sigma}\!\cdot\!
            \Bigl(\mathbf{D}\!\times\!\mathbf{E}-\mathbf{E}\!\times\!\mathbf{D}\Bigr)
      \Bigr\}
\\[3mm]\hspace{12.5mm}\displaystyle
   \left.
   -\frac{3e}{64m^4}\,\Bigl\{\mathbf{D}^2,\left[\boldsymbol{\nabla},\mathbf{E}\right]\Bigr\}
   -\frac{5e}{128m^4}\left[\mathbf{D}^2,(\mathbf{D}\!\cdot\!\mathbf{E}+\mathbf{E}\!\cdot\!\mathbf{D})\right]
   -\frac{e^2}{8m^3}\mathbf{E}^2\right)\psi_e,
\end{array}
\end{equation}
where $\mathbf{D}=\boldsymbol{\nabla}-ie\mathbf{A}$. The contact terms required in the NRQED theory~\cite{Caswell86,Kinoshita96,Hill13} are not considered here, because they do not play any role in the spin-orbit and spin-spin tensor interactions which are our main focus in the following. Here and in what follows we use the notation: $\{X,Y\}=XY+Y^*X^*$, $[X,Y]=XY-YX$ where the star denotes a Hermitian conjugate. The coupling constants, $c_i$, are determined by requiring that scattering amplitudes in QED and NRQED agree up to a chosen order in $\alpha$ and in $v^2/c^2$. Performing this matching at tree level, which is enough for the work presented here, one gets $c_F=c_D=c_S=c_{W_1}=1$.

As shown in more detail in~\cite{Haidar20}, the effective Hamiltonian $H_{\rm eff}$, which stems from the Lagrangian, is equivalent to the Foldy-Wouthuysen Hamiltonian $H_{\rm FW}$ derived in \cite{Patkos16} (see Eq.(23)). It may be obtained from $H_{\rm FW}$ through the canonical transformation $e^{iS}(H-i\partial_t)e^{-iS}$, where \cite{Patkos16}
\[
S = \frac{e}{8m^2} \, \boldsymbol{\sigma}\!\cdot\!\bigl(\mathbf{A}\!\times\!\boldsymbol{\pi}
                                             -\boldsymbol{\pi}\!\times\!\mathbf{A}\bigr),
\qquad
\boldsymbol{\pi} = \mathbf{p}-e\mathbf{A},
\]

where $\mathbf{p} = -i\boldsymbol{\nabla}$ is the electron's impulse. Heavy particles of masses $M_a$ charges $Z_a$, and impulses $\mathbf{P}_a$ with $a=1,2$, are treated within the leading-order interaction Hamiltonian:
\begin{equation}\label{eq:Hnuc}
H_I = -Z_a|e|\left(\frac{\mathbf{P}_a}{2M_a}\,\mathbf{A}+\mathbf{A}\,\frac{\mathbf{P}_a}{2M_a}\right) - \boldsymbol{\mu}_a\!\cdot\!\mathbf{B} + \frac{Z_a^2e^2}{2M_a}\mathbf{A}^2.
\end{equation}
The magnetic moments of particles are expressed as follows:
\[
\boldsymbol{\mu}_e=2\mu_e\mu_B\mathbf{s}_e=-\frac{(1+a_e)|e|}{m}\,\mathbf{s}_e,
\qquad
\boldsymbol{\mu}_a = \mu_a\,\mu_N\,\frac{\mathbf{I}}{I},
\qquad
\mu_N = \frac{|e|}{2m_p},
\]
$\mu_e$ and $\mu_a$ are dimensionless quantities measured in Bohr and nuclear magnetons, respectively.

We will consider corrections to the bound states of a one-electron molecular system such as H$_2^+$ or HD$^+$. The zero-order approximation is the nonrelativistic Schr\"odinger equation with the Hamiltonian
\begin{equation}\label{Hamiltonian}
H_0 = \frac{\mathbf{P}_1^2}{2M_1}+\frac{\mathbf{P}_2^2}{2M_2}+\frac{\mathbf{p}_e^2}{2m} + V, \hspace{3mm} V =
      -\frac{Z_1\alpha}{r_1}-\frac{Z_2\alpha}{r_2}+\frac{Z_1Z_2\alpha}{R}.
\end{equation}
Here $\mathbf{r}_a = \mathbf{r}_e\!-\!\mathbf{R}_a$ with $a=(1,2)$ is the electron's position with respect to the nucleus $a$, and $\mathbf{R}=\mathbf{R_2}\!-\!\mathbf{R}_1$ the internuclear vector. It is assumed that $M_a \gg m$. We also assume that the Hamiltonian is written in the center of mass (center of inertia) frame, which implies: $\mathbf{p}_e\!+\!\mathbf{P}_1\!+\!\mathbf{P}_2=0$.

The potentials $A_0$ and $\mathbf{A}$ are related to electric and magnetic field strengths as follows
\[
\mathbf{E} = -\boldsymbol{\nabla}A_0-\frac{\partial\mathbf{A}}{\partial t},
\qquad
\mathbf{B} = \boldsymbol{\nabla}\times\mathbf{A}.
\]
We define $\mathbf{E}_{\parallel}=-\boldsymbol{\nabla}A_0$ and $\mathbf{E}_{\perp}=-\frac{\partial\mathbf{A}}{\partial t}$, while $\mathbf{B}$ is always transverse. It is worth noting that $\mathbf{E}_{\parallel}$ corresponds to an instantaneous interaction, while $\mathbf{A}$ propagates in time with the velocity of light.

In order to determine which terms are needed at a given order, it is useful to know the nominal order of expectation values of various operators for a wavefunction of the nonrelativistic bound system. One gets~\cite{Kinoshita96}:
\[
\begin{array}{@{}l}
\left\langle \mathbf{p} \right\rangle \sim m(v/c),
\qquad
\left\langle \partial_t \right\rangle \sim m(v/c)^2,
\qquad
\left\langle eA_0 \right\rangle \sim m(v/c)^2,
\\[3mm]
\left\langle e\mathbf{A} \right\rangle \sim m(v/c)^3,
\qquad
\left\langle e\mathbf{E}_{\parallel} \right\rangle \sim m^2(v/c)^3,
\qquad
\left\langle e\mathbf{B} \right\rangle \sim m^2(v/c)^4,
\end{array}
\]
where $v$ is the typical velocity of the bound electron.

The photon propagator in the Coulomb gauge is:
\begin{equation}\label{C_gauge}
G^{\mu\nu} =
\begin{cases}
\displaystyle
G^{00} = \frac{1}{\mathbf{q}^2}\>,
                 & \mbox{--- the Coulomb photon propagator,}
\\[3mm]
\displaystyle
G^{ij} = \frac{\delta_{ij}-q_iq_j/\mathbf{q}^2}{q_0^2-\mathbf{q}^2+i\varepsilon}\>,
       & \mbox{--- the transverse photon propagator. }\\[3mm]
\end{cases}
\end{equation}

We use Feynman's time-ordered perturbation formalism~\cite{Feynman49}, whereby the change of energy of a bound system due to exchange of one photon is expressed
\begin{equation}\label{perturb_E}
\begin{array}{@{}l}\displaystyle
\Delta E = \int\frac{d^4q}{(2\pi)^4i}
     G^{\mu\nu}(q)
     \left\langle
        \psi_0\left|
           V_\mu(2)\,e^{i\mathbf{qr}_a}\,
           \frac{1}{E_0\!-\!q_0\!-\!H_0}
           \,e^{-i\mathbf{qr}_b}\,V_\nu(1)
        \right|\psi_0
     \right\rangle
\end{array}
\end{equation}
where $V(1)$ and $V(2)$ are some NRQED vertices for the electron or nucleus, $\psi_0$ and $E_0$ are respectively the nonrelativistic bound state wave function and energy for the Hamiltonian (\ref{Hamiltonian}), and $\mathbf{r}_a,\mathbf{r}_b$ the position operators of the involved particles. This formula dates back from the original work by Feynman~\cite{Feynman53} (Sec.\ II) and appears in a slightly modified form in~\cite{Pachucki04,Pachucki05}.


\section{NRQED vertices} \label{vertices}

It is convenient to translate the NRQED Lagrangian [Eq.~(\ref{Lagrangian})] in terms of NRQED vertices and ``Feynman'' rules, as done in Fig.~3 of Ref.~\cite{Kinoshita96}. Here, we list the vertices contributing to the $m\alpha^6$ and $m\alpha^6(m/M)$ orders, and give their expressions both in momentum and coordinate space, which are connected to each other by a 3D Fourier transformation. In the momentum space we use $\mathbf{p}$ and $\mathbf{p}'$ as momenta of incident and scattered electron, $\mathbf{q}=\mathbf{p}'\!-\!\mathbf{p}$ is the transferred momentum.

We first give the tree-level vertices related to the electron line:
\begin{equation}\label{vertex:tree}
\begin{array}{@{}l@{\hspace{5mm}}l@{\hspace{8mm}}l}
1. & \displaystyle
     \phantom{-}e\left(\frac{3\mathbf{q}^2(p'^2\!+\!p^2)}{64m^4}+\frac{5(p'^2\!-\!p^2)^2}{128m^4}\right)A_0
   & \displaystyle
     -\frac{3e}{64m^4}\Bigl\{p^2,\left[\Delta A_0\right]\Bigr\}+\frac{5e}{128m^4}\left[p^2,[p^2,A_0]\right]
\\[3mm]
2. & \displaystyle -e\left(i\frac{3\boldsymbol{\sigma}[\mathbf{q}\times\mathbf{p}](p'^2\!+\!p^2)}{32m^4}\right)A_0
   & \displaystyle
     \phantom{-}
     \frac{3e}{32m^4}\,\Bigl\{p^2,\boldsymbol{\sigma}\!\cdot\![\mathbf{E}_{\parallel}\!\times\!\mathbf{p}]\Bigr\}
\\[3mm]
3. & \displaystyle
     \phantom{-}e\frac{(p'^2\!+\!p^2)(\mathbf{p}'\!+\!\mathbf{p})}{8m^3}\,\mathbf{A}
   & \displaystyle
     \phantom{-}\frac{e}{8m^3}\,\Bigl\{p^2,\mathbf{p}\!\cdot\!\mathbf{A}\!+\!\mathbf{A}\!\cdot\!\mathbf{p}\Bigr\}
\\[3mm]
4. & \displaystyle \phantom{-}e\frac{i[\boldsymbol{\sigma}\!\times\!\mathbf{q}](p'^2\!+\!p^2)}{8m^3}\,\mathbf{A}
   & \displaystyle
     \phantom{-}\frac{e}{8m^3}\,\Bigl\{p^2,\boldsymbol{\sigma}\!\cdot\!\mathbf{B}\Bigr\}
\\[3mm]
5. & \displaystyle -e\frac{\mathbf{p'}\!+\!\mathbf{p}}{2m}\mathbf{A}
             & \displaystyle -e\left(\frac{\mathbf{p}}{2m}\,\mathbf{A}+\mathbf{A}\,\frac{\mathbf{p}}{2m}\right)
\\[3mm]
6. & \displaystyle -e\frac{i}{2m}\left[\boldsymbol{\sigma}\!\times\!\mathbf{q}\right]\!\cdot\!\mathbf{A}
             & \displaystyle -\frac{e}{2m} \, \boldsymbol{\sigma}\!\cdot\!\mathbf{B}
\\[3mm]
7. & \displaystyle
     \phantom{-}e\frac{iq_0[\boldsymbol{\sigma}\!\times\!(\mathbf{p}'\!+\!\mathbf{p})]}{8m^2}\,\mathbf{A}
   & \displaystyle
     -\frac{e}{8m^2}\,\boldsymbol{\sigma}\!\cdot\!
        \Bigl(\mathbf{p}\!\times\!\partial_t\mathbf{A}-\partial_t\mathbf{A}\!\times\!\mathbf{p}\Bigr)
\end{array}
\end{equation}
The last one appears only in the retardation contribution, see Sec~\ref{sec-timederiv}. For nuclei, the following tree-level vertices come into play:
\begin{equation} \label{vertex:nuclear}
\begin{array}{@{}l@{\hspace{8mm}}l@{\hspace{33mm}}l}
1\textrm{N}. & \displaystyle  \phantom{-}Z_a |e|A_0
             & \displaystyle  \phantom{-}Z_a |e|A_0
\\[3mm]
2\textrm{N}. & \displaystyle -Z_a |e|\frac{\mathbf{P}_a\!+\!\mathbf{P}'_a}{2M_a}\mathbf{A}
             & \displaystyle -Z_a |e|\left(\frac{\mathbf{P}_a}{2M_a}\,\mathbf{A}+\mathbf{A}\,\frac{\mathbf{P}_a}{2M_a}\right)
\\[3mm]
3\textrm{N}. & \phantom{-}\displaystyle i\left[\boldsymbol{\mu}_a\!\times\!\mathbf{q}\right]\!\cdot\!\mathbf{A}
             & -\boldsymbol{\mu}_a\!\cdot\!\mathbf{B}
\end{array}
\end{equation}

The seagull-type vertices for the electron are:
\begin{equation}\label{eq:seagull}
\begin{array}{@{}l@{\hspace{9mm}}l@{\hspace{22mm}}l}
8. & \displaystyle
   \phantom{-}e^2\frac{i\mathbf{q}_1\!\times\!\boldsymbol{\sigma}}{4m^2}A_0(\mathbf{q}_1)\mathbf{A}(\mathbf{q}_2)
   & \displaystyle
   -\frac{e^2}{8m^2}\boldsymbol{\sigma}\Bigl(\mathbf{A}\!\times\!\mathbf{E}-\mathbf{E}\!\times\!\mathbf{A}\Bigr)
\\[3mm]
9. & \displaystyle -e^2\frac{q_1^iq_2^i}{8m^3}A_0(\mathbf{q}_1)A_0(\mathbf{q}_2)
   & \phantom{-}\displaystyle \frac{e^2}{8m^3}\mathbf{E}^2
\\[3mm]
10. & \displaystyle \phantom{-} e^2\frac{\delta_{ij}}{2m}\mathbf{A}(\mathbf{q}_1)\mathbf{A}(\mathbf{q}_2)
   & \phantom{-}\displaystyle \frac{e^2}{2m}\mathbf{A}^2
\end{array}
\end{equation}
Note that two transverse photon vertex 10. only contributes at the $(m/M)^2$ order and thus will not be used in the following. However, the corresponding vertex for nuclei should be included:
\begin{equation}\label{eq:Vnuc}
4\textrm{N}.  \hspace{9mm} Z_a^2e^2\frac{\delta_{ij}}{2M_a}\mathbf{A}(\mathbf{q}_1)\!\cdot\!\mathbf{A}(\mathbf{q}_2)
\hspace{22mm}
\phantom{-}\displaystyle \frac{Z_a^2e^2}{2M_a}\mathbf{A}^2.
\end{equation}

In the following, we obtain from these vertices the effective potentials at orders $m\alpha^6$ and $m\alpha^6(m/M)$ (both spin-independent and spin-dependent) by systematic application of the nonrelativistic Rayleigh-Schr\"odinger perturbation theory. For each term, we will mention which vertices are involved by referring to the numbering given above. It is understood that all terms should be summed over the nuclear index $a$ ($a=1,2$, and $b=3-a$).

\section{Tree-level interactions} \label{tree}

We first consider the tree-level diagrams involving the exchange of one photon between the electron and a nucleus. The derivation of effective potentials is straightforward in this case (one such example is given in~\cite{Haidar20}). For the transformation from momentum to coordinate space, useful integrals can be found in Appendix~\ref{Fourier}. Two terms come from a Coulomb photon exchange (vertices 1-1N and 2-1N):

\begin{equation}
\begin{array}{@{}l}\displaystyle
\mathcal{U}_{1a} =
   -\frac{3}{64m^4}\,
         \left\{p_e^2,\left[\Delta V\right]\right\}
   +\frac{5}{128m^4}\left[p_e^2,[p_e^2,V]\right]
\\[3mm]\displaystyle
\mathcal{U}_{1b} 
 = -\frac{3Z_a\alpha}{32m^4}\,
         \left\{p_e^2,\frac{1}{r_a^3}[\mathbf{r}_a\!\times\!\mathbf{p}_e]\!\cdot\!\boldsymbol{\sigma}_e\right\}.
\end{array}
\end{equation}
The square brackets around quantities imply that derivatives act only within the bracket, thus $[\Delta V]$ in the first line corresponds to the Laplacian of the Coulomb potential i.e. a sum of delta-function operators. The transverse photon exchange produces four terms (3-2N, 3-3N, 4-2N and 4-3N):
\begin{equation} \label{tr-ph}
\begin{array}{@{}l}\displaystyle
\mathcal{U}_{2a} =
   -\frac{Z_a\alpha}{8m^2}\>
      \left\{p_e^2,\frac{p_e^i}{m}\left(\frac{\delta^{ij}}{r_a}+\frac{r_a^ir_a^j}{r_a^3}\right)\frac{P_a^j}{M_a}\right\}
\\[4mm]\displaystyle
\mathcal{U}_{2b} =
   \phantom{-}\frac{Z_a\alpha}{8m^3M_a}\,
   \left\{p_e^2,\frac{1}{r_a^3}\bigl[\mathbf{r}_a\!\times\!\mathbf{P}_a\bigr]\!\cdot\!\boldsymbol{\sigma}_e\right\}
\\[4mm]\displaystyle
\mathcal{U}_{2c} =
   -\frac{\alpha}{4m^3}\,
      \left\{p_e^2,\frac{1}{r_a^3}[\mathbf{r}_a\!\times\!\mathbf{p}_e]\,\frac{\boldsymbol{\mu}_a}{|e|}\right\}
\\[4mm]\displaystyle
\mathcal{U}_{2d} =
   \phantom{-}\frac{1}{4m^2}\,
      \left\{p_e^2,
      \left[
            \frac{8\pi}{3}\boldsymbol{\mu}_e\boldsymbol{\mu}_a\,\delta^3(\mathbf{r}_a)
            -\frac{r_a^2\boldsymbol{\mu}_e\boldsymbol{\mu}_a
                 \!-\!3(\boldsymbol{\mu}_e\mathbf{r}_a)
                       (\boldsymbol{\mu}_a\mathbf{r}_a)}{r_a^5}
      \right]\right\}.
\end{array}
\end{equation}

\section{Retardation in the single transverse photon exchange} \label{sec-retard}

According to Eq.~(\ref{perturb_E}), the energy correction due to a single transverse photon exchange between the electron and a nucleus is
\begin{equation}\label{retard}
\begin{array}{@{}l}\displaystyle
\Delta E = \frac{1}{(2\pi)^4}\int\frac{d^4q}{q^2\!+\!i\epsilon}
     \left(\delta_{ij}-\frac{q_iq_j}{\mathbf{q}^2}\right)
     \left\langle
        \psi_0\left|
           V^i(2)\,e^{i\mathbf{qr}_e}\,
           \frac{1}{E_0\!-\!q_0\!-\!H_0}
           \,e^{-i\mathbf{qR}_a}\,V^j(1)
        \right|\psi_0
     \right\rangle
\end{array}
\end{equation}
where $V(1)$ and $V(2)$ are some NRQED vertices from Section III, Eqs.~(\ref{vertex:tree}) and (\ref{vertex:nuclear}).

\subsection{Dipole and Fermi vertices}

Let us consider first the contribution from the leading-order vertices: 5 and 6 for the electron, 2N and 3N for the nucleus,
\begin{equation}\label{eq:retard}
\begin{array}{@{}l}
\displaystyle
\mathcal{U}_3^{(5+)} =
   -\frac{e}{(2\pi)^3}\int\frac{d\mathbf{q}}{2q}
     \left(\delta^{ij}\!-\!\frac{q^iq^j}{\mathbf{q}^2}\right)
     \Biggl\{
        \left(\frac{\mathbf{p}_e}{m}+\frac{i[\boldsymbol{\sigma}_e\!\times\!\mathbf{q}]}{2m}\right)^i
        \biggl[
           e^{i\mathbf{qr}_e}\left(\frac{1}{E_0\!-\!q\!-\!H_0}\!+\!\frac{1}{q}\right)e^{-i\mathbf{qR}_a}
\\[4mm]\displaystyle\hspace{25mm}
           -\left(\frac{1}{E_0\!-\!q\!-\!H_0}\!+\!\frac{1}{q}\right)
        \biggr]
        \left(-Z_a|e|\frac{\mathbf{P}_a}{M_a}+i\left[\boldsymbol{\mu}_a\!\times\!\mathbf{q}\right]\right)^j
     \Biggr\}
     +\left(
        e^{i\mathbf{qr}_e}\leftrightarrow e^{-i\mathbf{qr}_e}
        \atop
        e^{-i\mathbf{qR}_a}\leftrightarrow e^{i\mathbf{qR}_a}
     \right),
\end{array}
\end{equation}
where $^{(5+)}$ means orders $m\alpha^5$ and higher. The term $1/q$ in parentheses of~(\ref{eq:retard}) corresponds to the subtracted leading $m\alpha^4$ order contribution to the Breit--Pauli interaction~\cite{Pachucki98}, and the $m\alpha^5$-order is removed by subtracting the term corresponding to the $\mathbf{q=0}$ limit. We use the retardation expansion:
\begin{equation} \label{expansion}
\frac{1}{E_0\!-\!q\!-\!H_0}\!+\!\frac{1}{q} =
   \frac{H_0\!-\!E_0}{q^2}-\frac{(H_0\!-\!E_0)^2}{q^3}+\dots
\end{equation}
for transverse photon momenta $q\!\sim\!(v/c)\gg(H_0\!-\!E_0)\!\sim\!(v/c)^2$ (the contribution from smaller momenta is suppressed after the performed subtractions). Here, the first term corresponds to a contribution of order $m\alpha^5$~\cite{Pachucki98}, and the second term contributes to order $m\alpha^6$. Then
\begin{equation}
\begin{array}{@{}l} \label{U3}
\displaystyle
\mathcal{U}_3^{(6)} =
   \frac{e}{(2\pi)^3}\int\frac{d\mathbf{q}}{2q^4}
     \left(\delta^{ij}\!-\!\frac{q^iq^j}{\mathbf{q}^2}\right)
     \Biggl\{
        \left(\frac{\mathbf{p}_e}{m}+\frac{i[\boldsymbol{\sigma}_e\!\times\!\mathbf{q}]}{2m}\right)^i
\\[4mm]\displaystyle\hspace{10mm}
        \times\left[
           e^{i\mathbf{qr}_e}\left(H_0\!-\!E_0\right)^2e^{-i\mathbf{qR}_a}
           -\left(H_0\!-\!E_0\right)^2
        \right]
        \left(-Z_a|e|\frac{\mathbf{P}_a}{M_a}+i\left[\boldsymbol{\mu}_a\!\times\!\mathbf{q}\right]\right)^j
     \Biggr\}
     +\left(
        e^{i\mathbf{qr}_e}\leftrightarrow e^{-i\mathbf{qr}_e}
        \atop
        e^{-i\mathbf{qR}_a}\leftrightarrow e^{i\mathbf{qR}_a}
     \right)
\end{array}
\end{equation}
From the relationship (with $a=1,2$ and $b=3-a$)
\[
\mathbf{R}_a = -\frac{m}{M} \mathbf{r}_a \mp \frac{M_b}{M} \mathbf{R},
\]
where $M=M_1+M_2+m$, and $\mp$ means a minus sign for $a=1$ and plus for $a=2$, one gets:
\[
\begin{array}{@{}l@{\hspace{10mm}}l}\displaystyle
\left[H_0,e^{-i\mathbf{qR}_a}\right] =
   e^{-i\mathbf{qR}_a}E_0\>\mathcal{O}\left(\frac{m}{M}\right).
\end{array}
\]
As a result,
\begin{equation}\label{eq:relation}
\begin{array}{@{}l}\displaystyle
e^{i\mathbf{qr}_e}\left(H_0\!-\!E_0\right)^2e^{-i\mathbf{qR}_a} = e^{i\mathbf{qr}_a}\left(H_0\!-\!E_0\right)^2
   +e^{i\mathbf{qr}_e}\left[H_0,e^{-i\mathbf{qR}_a}\right]\left(H_0\!-\!E_0\right)
   +e^{i\mathbf{qr}_e}\left(H_0\!-\!E_0\right)\left[H_0,e^{-i\mathbf{qR}_a}\right]
\\[3mm]\hspace{36mm}\displaystyle
 \approx
   e^{i\mathbf{qr}_a}\left(H_0\!-\!E_0\right)^2
 = \left(H_0\!-\!E_0\right)e^{i\mathbf{qr}_a}\left(H_0\!-\!E_0\right)
   +\left[e^{i\mathbf{qr}_a},H_0\right]\left(H_0\!-\!E_0\right).
\end{array}
\end{equation}
In the second line, we have kept only the leading-order term in $(m/M)$.

Using this relationship, one immediately sees that the terms of Eq.~(\ref{U3}) involving the nuclear magnetic moment give a zero contribution when applied to the zero-order state $|\psi_0\rangle$. These terms thus contribute only at higher orders in $m/M$ ($m\alpha^6(m/M)^2$ and above) and will not be considered here. The remaining terms can be separated into a spin-independent term and a term contributing to the spin-orbit interaction.

For the spin-independent part we have
\[
\begin{array}{@{}l}
\displaystyle
\mathcal{U}_{3a} =
   \frac{Z_ae^2}{mM_a}\frac{1}{(2\pi)^3}\int\frac{d\mathbf{q}}{2q^4}
     \left(\delta^{ij}\!-\!\frac{q^iq^j}{\mathbf{q}^2}\right)p_e^i
     \Bigl\{
            \left(H_0\!-\!E_0\right)\left(e^{i\mathbf{qr}_a}-1\right)\left(H_0\!-\!E_0\right)
           +\left[e^{i\mathbf{qr}_a}-1,H_0\right]\left(H_0\!-\!E_0\right)
\\[3mm]\displaystyle\hspace{53mm}
           +\left(H_0\!-\!E_0\right)\left(e^{-i\mathbf{qr}_a}-1\right)\left(H_0\!-\!E_0\right)
           +\left[e^{-i\mathbf{qr}_a}-1,H_0\right]\left(H_0\!-\!E_0\right)
     \Bigr\}P_a^j,
\end{array}
\]
and after integration we finally obtain (using the third line of Eq.~(\ref{ap:Fourier})):
\begin{equation}
\begin{array}{@{}l}\displaystyle \label{U3a}
\mathcal{U}_{3a} =
   \frac{Z_a\alpha}{16mM_a}
   \left\{
      \left[p_e^i,V\right]\frac{r_a^ir_a^j-3r_a^2\delta^{ij}}{r_a}\left[V,P_a^j\right]
      +p_e^i\left[\frac{r_a^ir_a^j-3r_a^2\delta^{ij}}{r_a},\frac{p_e^2}{2m}\right]\left[V,P_a^j\right]
   \right\}+(h.c.)
\\[3mm]\hspace{6mm}\displaystyle
 = \frac{Z_a^2\alpha^3}{8mM_a}
      \left[\frac{Z_1r^i_1}{r_1^3}+\frac{Z_2r_2^i}{r_2^3}\right]\frac{r_a^ir_a^j-3r_a^2\delta^{ij}}{r_a}
           \left[-\frac{r^j_a}{r_a^3}\pm\frac{Z_bR^j}{R^3}\right]
\\[3mm]\hspace{10mm}\displaystyle
      +\frac{Z_a^2\alpha^2}{8m^2M_a}
         \left[
            \frac{p_e^2}{r_a^2}-3\frac{\mathbf{r}_a(\mathbf{r}_a\mathbf{p}_e)\mathbf{p}_e}{r_a}
         \right]
      \mp\frac{Z_a^2Z_b\alpha^2}{8m^2M_a}
         \left[
            \frac{(\mathbf{Rr}_a)p_e^2}{R^3r_a}
            -\frac{(\mathbf{Rr}_a)\mathbf{r}_a(\mathbf{r}_a\mathbf{p}_e)\mathbf{p}_e}{R^3r_a^3}
            -2\frac{\mathbf{R}(\mathbf{r}_a\mathbf{p}_e)\mathbf{p}_e}{R^3r_a}
         \right].
\end{array}
\end{equation}
Here we have used
\[
[V,\mathbf{P}_a] = i\left(-\frac{Z_a\alpha \, \mathbf{r}_a}{r_a^3} \pm \frac{Z_a Z_b \alpha \mathbf{R}}{R^3}\right).
\]
Contributions from the $\mathbf{R}/R^3$ terms can be assumed to be small, since at small $R$ the wave function is exponentially small due to the strong Coulomb barrier. Then the interaction may be simplified to
\begin{equation}
\begin{array}{@{}l}\displaystyle
\mathcal{U}_{3a} =
   \frac{\alpha^3}{4m}
      \left[
         \frac{Z_1^3}{M_1}\frac{1}{r_1^3}+\frac{Z_2^3}{M_2}\frac{1}{r_2^3}
         +\frac{Z_1^2Z_2}{M_1}\frac{(\mathbf{r}_1\mathbf{r}_2)}{r_1^2r_2^3}
         +\frac{Z_1Z_2^2}{M_2}\frac{(\mathbf{r}_1\mathbf{r}_2)}{r_1^3r_2^2}
      \right]
\\[3mm]\hspace{8mm}\displaystyle
      +\frac{Z_1^2\alpha^2}{8m^2M_1}
         \left[
            \frac{1}{r_1^4}+\mathbf{p}_e\frac{1}{r_1^2}\mathbf{p}_e
               -3\frac{(\mathbf{p}_e\mathbf{r}_1)(\mathbf{r}_1\mathbf{p}_e)}{r_1^4}
         \right]
      +\frac{Z_2^2\alpha^2}{8m^2M_2}
         \left[
            \frac{1}{r_2^4}+\mathbf{p}_e\frac{1}{r_2^2}\mathbf{p}_e
               -3\frac{(\mathbf{p}_e\mathbf{r}_2)(\mathbf{r}_2\mathbf{p}_e)}{r_2^4}
         \right].
\end{array}
\end{equation}

The electron spin-orbit term is
\[
\begin{array}{@{}l}
\displaystyle
\mathcal{U}_{3b} =
   \frac{Z_ae^2}{2mM_a}\frac{1}{(2\pi)^3}\int\frac{d\mathbf{q}}{2q^4}
     \>i[\boldsymbol{\sigma}_e\!\times\!\mathbf{q}]\,
     \Bigl\{
           \left[e^{i\mathbf{qr}_a}-1,H_0\right]\left(H_0\!-\!E_0\right)
           +\left[e^{-i\mathbf{qr}_a}-1,H_0\right]\left(H_0\!-\!E_0\right)
     \Bigr\}\mathbf{P}_a
\end{array}
\]
After Fourier transform:
\[
\begin{array}{@{}l}\displaystyle
\mathcal{U}_{3b} =
   -\frac{Z_a\alpha}{16m^2M_a}
      \left[p_e^2,\left[\frac{\mathbf{r}_a}{r_a}\!\times\!\boldsymbol{\sigma}_e\right]\right]
      \left[V,\mathbf{P}_a\right]+(h.c.),
   \end{array}
\]
and using the commutators
\[
\left[P_a^2,\frac{r_a^j}{r_a}\right] = \frac{2r_a^j}{r_a^3}-\frac{2i}{r_a}P_a^j+\frac{2ir_a^j}{r_a^3}\left(\mathbf{r}_a\mathbf{P}_a\right),
\qquad \left[\mathbf{P}_a\mathbf{P}_b,\frac{r_a^j}{r_a}\right] = -\frac{i}{r_a}P_b^j + \frac{ir_a^j}{r_a^3}\left(\mathbf{r}_a\mathbf{P}_b\right),
\]
one gets:
\begin{equation} \label{u3b}
\begin{array}{@{}l}\displaystyle
\mathcal{U}_{3b} = \frac{Z_a^2\alpha^2}{4m^2M_a}
    \left[
      \frac{1}{r_a^4}\left[\mathbf{r}_a\!\times\!\mathbf{p}_e\right]
      \mp\frac{Z_b}{r_aR^3}[\mathbf{R}\!\times\!\mathbf{p}_e]
      \mp\frac{Z_b}{r_a^3R^3}[\mathbf{r}_a\!\times\!\mathbf{R}](\mathbf{r}_a\mathbf{p}_e)
    \right]\!\cdot\!\boldsymbol{\sigma}_e
\\[3mm]\hspace{6mm}\displaystyle
 = \frac{Z_1^2\alpha^2}{4m^2M_1}\frac{1}{r_1^4}\left[\mathbf{r}_1\!\times\!\mathbf{p}_e\right]\!\cdot\!\boldsymbol{\sigma}_e
   +\frac{Z_2^2\alpha^2}{4m^2M_2}\frac{1}{r_2^4}\left[\mathbf{r}_2\!\times\!\mathbf{p}_e\right]\!\cdot\!\boldsymbol{\sigma}_e
\\[3mm]\hspace{10mm}\displaystyle
  -\frac{Z_1^2Z_2\alpha^2}{4m^2M_1}\,
    \frac{1}{r_1R^3}\,[\mathbf{R}\!\times\!\mathbf{p}_e]\!\cdot\!\boldsymbol{\sigma}_e
  +\frac{Z_2^2Z_1\alpha^2}{4m^2M_2}\,
    \frac{1}{r_2R^3}\,[\mathbf{R}\!\times\!\mathbf{p}_e]\!\cdot\!\boldsymbol{\sigma}_e
\\[3mm]\hspace{10mm}\displaystyle
  +\frac{Z_1^2Z_2\alpha^2}{4m^2M_1}\,
      \frac{[\mathbf{r}_1\!\times\!\mathbf{r}_2]}{r_1^3R^3}\,(\mathbf{r}_1\mathbf{p}_e)\!\cdot\!\boldsymbol{\sigma}_e
  -\frac{Z_2^2Z_1\alpha^2}{4m^2M_2}\,
      \frac{[\mathbf{r}_1\!\times\!\mathbf{r}_2]}{r_2^3R^3}\,(\mathbf{r}_2\mathbf{p}_e)\!\cdot\!\boldsymbol{\sigma}_e.
\end{array}
\end{equation}
Similarly to the $\mathcal{U}_{3a}$ term, one can neglect the last two lines in the above expression.

\subsection{Time derivative vertex} \label{sec-timederiv}

Now we consider a retardation term where the electron interacts via the time derivative vertex (number~7 in Eq.~(\ref{vertex:tree})) while the nucleus interacts via the lowest-order vertices (dipole (2N) or Fermi (3N)).
\begin{equation}\label{time_deriv}
\begin{array}{@{}l}\displaystyle
\mathcal{U}_{3c}^{(5+)} = \frac{(-i)}{(2\pi)^4}\int\frac{d^4q}{q_0^2\!-\!\mathbf{q}^2\!+\!i\epsilon}
     \left(\delta_{ij}-\frac{q_iq_j}{\mathbf{q}^2}\right)
        \left[\frac{ieq^0}{8m^2}\,(\mathbf{p}_e\!+\!\mathbf{p}'_e)\!\times\!\boldsymbol{\sigma}_e\right]^i\times
\\[3mm]\displaystyle\hspace{20mm}
        \times\left\{
           \,e^{i\mathbf{qr}_e}\,
           \frac{1}{E_0\!-\!q_0\!-\!H_0\!+\!i\epsilon}
           \,e^{-i\mathbf{qR}_a}\,
        \right\}
        \left(-Z_ae\frac{\mathbf{P}_a}{M_a}+i\left[\boldsymbol{\mu}_a\!\times\!\mathbf{q}\right]\right)^j
     +\left(
        e^{i\mathbf{qr}_e}\leftrightarrow e^{-i\mathbf{qr}_e}
        \atop
        e^{-i\mathbf{qR}}\leftrightarrow e^{i\mathbf{qR}}
     \right),
\end{array}
\end{equation}
The first step is integration over $q_0$. Using integration in the complex plane one gets
\begin{equation} \label{timederiv}
\begin{array}{@{}l}\displaystyle
\mathcal{U}_{3c}^{(5+)} = \frac{ie}{16m^2}\int\frac{d^3q}{(2\pi)^3}
     \left(\delta_{ij}-\frac{q_iq_j}{\mathbf{q}^2}\right)
        \left[(\mathbf{p}_e\!+\!\mathbf{p}'_e)\!\times\!\boldsymbol{\sigma}_e\right]^i\times
\\[3mm]\displaystyle\hspace{20mm}
        \times\left\{
           \,e^{i\mathbf{qr}_e}\,
           \frac{1}{E_0\!-\!q\!-\!H_0}
           \,e^{-i\mathbf{qR}_a}\,
        \right\}
        \left(-Z_ae\frac{\mathbf{P}_a}{M_a}+i\left[\boldsymbol{\mu}_a\!\times\!\mathbf{q}\right]\right)^j
     +\left(
        e^{i\mathbf{qr}_e}\leftrightarrow e^{-i\mathbf{qr}_e}
        \atop
        e^{-i\mathbf{qR}_a}\leftrightarrow e^{i\mathbf{qR}_a}
     \right),
\end{array}
\end{equation}
The time derivative vertex is of nominal order $(v/c)^3 \sim \alpha^3$. The first term in the expansion of $1/(E_0 - H_0 - q)$ (i.e. $-1/q$, see Eq.~(\ref{expansion})) would produce a contribution of order $m\alpha^5$, but this contribution vanishes due to cancellation between both terms of Eq.~(\ref{timederiv}). The $m\alpha^6$-order term corresponds to the next term, $(H_0-E_0)/q^2$. Using the relation $e^{i\mathbf{qr}_e}\left(H_0\!-\!E_0\right)e^{-i\mathbf{qR}_a} \approx e^{i\mathbf{qr}_a}\left(H_0\!-\!E_0\right)$, similar to Eq.~(\ref{eq:relation}), one gets that the term involving the nuclear magnetic moment only contributes at the $(m/M)^2$ order and mat thus be ignored. The remaining term is
\begin{equation}
\begin{array}{@{}l}\displaystyle
\mathcal{U}_{3c}^{(6)}
   \approx -\frac{iZ_ae^2}{8M_am^2}\int\frac{d^3q}{(2\pi)^3}
     \frac{1}{\mathbf{q}^2}\left(\delta_{ij}-\frac{q_iq_j}{\mathbf{q}^2}\right) \times
\\[3mm]\displaystyle\hspace{25mm}
     \times\left\{
        \left[\mathbf{p}_e\!\times\!\boldsymbol{\sigma}_e\right]^i
        e^{i\mathbf{qr}_a}(H_0\!-\!E_0)P_a^j
        +\left[\mathbf{p}_e\!\times\!\boldsymbol{\sigma}_e\right]^i
        e^{-i\mathbf{qr}_a}(H_0\!-\!E_0)P_a^j
     \right\}
\\[3mm]\displaystyle\hspace{6.5mm}
   = -\frac{iZ_ae^2}{8M_am^2}\int\frac{d^3q}{(2\pi)^3}
     \frac{1}{\mathbf{q}^2}\left(\delta_{ij}-\frac{q_iq_j}{\mathbf{q}^2}\right) \times
\\[3mm]\displaystyle\hspace{25mm}
     \times\left\{
        \left[\mathbf{p}_e\!\times\!\boldsymbol{\sigma}_e\right]^i
        \left(e^{i\mathbf{qr}_a}\!-\!1\right)[H_0,P_a^j]
        +\left[\mathbf{p}_e\!\times\!\boldsymbol{\sigma}_e\right]^i
        \left(e^{-i\mathbf{qr}_a}\!-\!1\right)[H_0,P_a^j]
     \right\}.
\end{array}
\end{equation}

After Fourier transform:
\begin{equation}\label{u3c}
\begin{array}{@{}l}\displaystyle
\mathcal{U}_{3c} =
   -\frac{iZ_a\alpha}{8M_am^2}\,
      [\mathbf{p}_e\!\times\!\boldsymbol{\sigma}_e]^i
      \>\frac{1}{2r_a}\left(\delta^{ij}+\frac{r_a^ir_a^j}{r_a^2}\right)
      \left[V,P_a^j\right]
   + (h.c)
\\[3mm]\displaystyle\hspace{6.5mm}
   = -\frac{Z_1^2\alpha^2}{4M_1m^2}\,\frac{1}{r_1^4}[\mathbf{r}_1\!\times\!\mathbf{p}_e]\!\cdot\!\boldsymbol{\sigma}_e
     -\frac{Z_2^2\alpha^2}{4M_2m^2}\,\frac{1}{r_2^4}[\mathbf{r}_2\!\times\!\mathbf{p}_e]\!\cdot\!\boldsymbol{\sigma}_e
\\[3mm]\displaystyle\hspace{10.5mm}
   + \frac{Z_1^2Z_2\alpha^2}{8M_1m^2}\,
         \frac{1}{r_1R^3}[\mathbf{R}\!\times\!\mathbf{p}_e]\!\cdot\!\boldsymbol{\sigma}_e
   - \frac{Z_2^2Z_1\alpha^2}{8M_2m^2}\,
      \frac{1}{r_2R^3}[\mathbf{R}\!\times\!\mathbf{p}_e]\!\cdot\!\boldsymbol{\sigma}_e
\\[3mm]\displaystyle\hspace{10.5mm}
   + \frac{Z_1^2Z_2\alpha^2}{8M_1m^2}\,
      \frac{(\mathbf{r}_1\mathbf{R})}{r_1^3R^3}[\mathbf{r}_1\!\times\!\mathbf{p}_e]\!\cdot\!\boldsymbol{\sigma}_e
   - \frac{Z_2^2Z_1\alpha^2}{8M_2m^2}\,
      \frac{(\mathbf{r}_2\mathbf{R})}{r_2^3R^3}[\mathbf{r}_2\!\times\!\mathbf{p}_e]\!\cdot\!\boldsymbol{\sigma}_e.
\end{array}
\end{equation}
Once more, the last two lines in the above expression may be neglected. Comparing Eq.~(\ref{u3b}) with Eq.~(\ref{u3c}), we see that the leading terms cancel out, so that $\mathcal{U}_{3b} + \mathcal{U}_{3c}$ is negligibly small. We will thus ignore these two terms when writing the total effective Hamiltonian in Sec.~\ref{heff6}.

\section{Seagull-type interactions} \label{seagull}

It remains to consider the contributions arising from seagull-type vertices, Eqs.~(\ref{eq:seagull}) and (\ref{eq:Vnuc}). All the interactions in momentum space are the convolution of two functions, which represent either the electric field strength $\mathbf{E}$ or magnetic field potential $\mathbf{A}$. In the coordinate space, they are directly given by a product (scalar or vector) of the same functions converted to the coordinate space.

The corresponding expressions for the electric field strength of a point-like Coulomb source, and for the magnetic field potential produced by the moving charge and magnetic moment of a nucleus, are
\[
\begin{array}{@{}l}\displaystyle
e\mathbf{E} = Z_a e^2\frac{i\mathbf{q}}{\mathbf{q}^2} \Rightarrow
   -Z_a \alpha\frac{\mathbf{r}_a}{r_a^3}
\\[3mm]\displaystyle
e\mathbf{A}_1 =
   -\frac{e}{\mathbf{q}^2}\left(\delta^{ij}-\frac{q^iq^j}{\mathbf{q}^2}\right)
      \left(-Z_a|e|\frac{\mathbf{P}_a}{M_a}\right) \Rightarrow
   -\frac{Z_a\alpha}{2M_a}\left(\frac{\delta^{ij}}{r_a}+\frac{r_a^ir_a^j}{r_a^3}\right)P_a^j
\\[3mm]\displaystyle
e\mathbf{A}_2 =
   \left[
      -\frac{e}{\mathbf{q}^2}
      \left(\delta^{ij}-\frac{q^iq^j}{\mathbf{q}^2}\right)
   \right]
   \Bigl(-i\left[\boldsymbol{\mu}_a\!\times\!(-\mathbf{q})\right]\Bigr) =
   \frac{ie}{\mathbf{q}^2}\left[\mathbf{q}\!\times\!\boldsymbol{\mu}_a\right] \Rightarrow
   -\frac{e}{r_a^3}\left[\mathbf{r}_a\!\times\!\boldsymbol{\mu}_a\right]
\end{array}
\]
and the potential produced by the electron at the location of a nucleus is
\[
|e|Z_a\mathbf{A}_a =
   \left[
      -\frac{|e|Z_a}{\mathbf{q}^2}
      \left(\delta^{ij}-\frac{q^iq^j}{\mathbf{q}^2}\right)
   \right]
   \biggl(
      -e\,\frac{\mathbf{p}_e}{m}
      +i\frac{e\left[\boldsymbol{\sigma}_e\!\times\!(-\mathbf{q})\right]}{2m}
   \biggr)
   \Rightarrow
   -\frac{Z_a\alpha}{2m}\left(\frac{\delta^{ij}}{r_a}+\frac{r_a^ir_a^j}{r_a^3}\right)p_e^j
   -\frac{Z_a\alpha}{2m}\>\frac{1}{r_a^3}\left[\mathbf{r}_a\!\times\!\boldsymbol{\sigma}_e\right]
\]

The first seagull-type contribution is a double Coulomb photon exchange diagram (vertices 7-1N-1N):
\begin{equation}
\mathcal{U}_{4} = \frac{e^2\mathbf{E}^2}{8m^3}
   = \frac{\alpha^2}{8m^3}\left[\frac{Z_1\mathbf{r}_1}{r_1^3}+\frac{Z_2\mathbf{r}_2}{r_2^3}\right]^2.
\end{equation}

The next terms stem from the seagull vertex with one Coulomb and one transverse photon (vertices 8-1N-2N and 8-1N-3N):
\[
\begin{array}{@{}l}\displaystyle
\mathcal{U}_{5a} =
     \frac{\boldsymbol{\sigma}_e}{4m^2}
     \left[
        \biggl[-Z_b\alpha\frac{r_b^i}{r_b^3}\biggr]
        \times
        \left\{
           -\frac{Z_a\alpha}{2M_a}\left(\frac{\delta^{ij}}{r_a}+\frac{r_a^ir_a^j}{r_a^3}\right)P_a^j
        \right\}
     \right]
\\[4mm]\displaystyle
\mathcal{U}_{5b} =
     \frac{\boldsymbol{\sigma}_e}{4m^2}
     \left[
        \biggl[-Z_b\alpha\frac{\mathbf{r}_b}{r_b^3}\biggr]
        \times
        \left(-\frac{e}{r_a^3}\left[\mathbf{r}_a\!\times\!\boldsymbol{\mu}_a\right]\right)
     \right] =
     -\frac{\alpha Z_b}{2m}\frac{1}{r_b^3}
     \left[\mathbf{r}_b\!\times\!\boldsymbol{\mu}_e\right]
        \frac{1}{r_a^3}\left[\mathbf{r}_a\!\times\!\boldsymbol{\mu}_a\right]
\end{array}
\]
or finally
\begin{equation}
\begin{array}{@{}l}\displaystyle
\mathcal{U}_{5a} =
   \frac{Z_1^2\alpha^2}{8m^2M_1}\>
     \frac{1}{r_1^4}\left[\mathbf{r}_1\!\times\!\mathbf{P}_1\right]\!\cdot\!\boldsymbol{\sigma}_e
   +\frac{Z_2^2\alpha^2}{8m^2M_2}\>
     \frac{1}{r_2^4}\left[\mathbf{r}_2\!\times\!\mathbf{P}_2\right]\!\cdot\!\boldsymbol{\sigma}_e
\\[3mm]\displaystyle\hspace{10mm}
   +\frac{Z_1Z_2\alpha^2}{8m^2M_1}\>
     \frac{1}{r_1r_2^3}\left[\mathbf{r}_2\!\times\!\mathbf{P}_1\right]\!\cdot\!\boldsymbol{\sigma}_e
   +\frac{Z_1Z_2\alpha^2}{8m^2M_2}\>
     \frac{1}{r_1^3r_2}\left[\mathbf{r}_1\!\times\!\mathbf{P}_2\right]\!\cdot\!\boldsymbol{\sigma}_e
\\[3mm]\displaystyle\hspace{10mm}
   -\frac{Z_1Z_2\alpha^2}{8m^2M_1}\>
     \frac{1}{r_1^3r_2^3}\left[\mathbf{r}_1\!\times\!\mathbf{r}_2\right](\mathbf{r}_1\mathbf{P}_1)\!\cdot\!\boldsymbol{\sigma}_e
   +\frac{Z_1Z_2\alpha^2}{8m^2M_2}\>
     \frac{1}{r_1^3r_2^3}\left[\mathbf{r}_1\!\times\!\mathbf{r}_2\right](\mathbf{r}_2\mathbf{P}_2)\!\cdot\!\boldsymbol{\sigma}_e
\\[4mm]\displaystyle
\mathcal{U}_{5b} =
    -\frac{\alpha}{6m}
    \left[
       Z_1\frac{r_1^2\boldsymbol{\mu}_e\boldsymbol{\mu}_1
             -3(\boldsymbol{\mu}_e\mathbf{r}_1)(\boldsymbol{\mu}_1\mathbf{r}_1)}{r_1^6}
       +Z_2\frac{r_2^2\boldsymbol{\mu}_e\boldsymbol{\mu}_2
             -3(\boldsymbol{\mu}_e\mathbf{r}_2)(\boldsymbol{\mu}_2\mathbf{r}_2)}{r_2^6}
       +\frac{2Z_1\boldsymbol{\mu}_e\boldsymbol{\mu}_1}{r_1^4}
       +\frac{2Z_2\boldsymbol{\mu}_e\boldsymbol{\mu}_2}{r_2^4}
    \right]
\\[3mm]\displaystyle\hspace{10mm}
    -\frac{\alpha}{6m}\,
    \left[
       Z_2\frac{(\mathbf{r}_1\mathbf{r}_2)\boldsymbol{\mu}_e\boldsymbol{\mu}_1
             -3(\boldsymbol{\mu}_e\mathbf{r}_1)(\boldsymbol{\mu}_1\mathbf{r}_2)}{r_1^3r_2^3}
       +Z_1\frac{(\mathbf{r}_1\mathbf{r}_2)\boldsymbol{\mu}_e\boldsymbol{\mu}_2
             -3(\boldsymbol{\mu}_e\mathbf{r}_2)(\boldsymbol{\mu}_2\mathbf{r}_1)}{r_1^3r_2^3}
    \right.
\\[3mm]\displaystyle\hspace{45mm}
    \left.
       +\frac{2Z_2(\mathbf{r}_1\mathbf{r}_2)\boldsymbol{\mu}_e\boldsymbol{\mu}_1}{r_1^3r_2^3}
       +\frac{2Z_1(\mathbf{r}_1\mathbf{r}_2)\boldsymbol{\mu}_e\boldsymbol{\mu}_2}{r_1^3r_2^3}
    \right]
\end{array}
\end{equation}

The last contribution to consider is a double transverse photon exchange with the top on a nucleus and two legs on an electron (vertices 4N-5-5, 4N-5-6, and 4N-6-6):
\begin{equation}
\begin{array}{@{}l}\displaystyle
\mathcal{U}_{6a} =
   \frac{Z_1^2\alpha^2}{8m^2M_1}
   \left[
      \mathbf{p}_e\frac{1}{r_1^2}\mathbf{p}_e+3\frac{(\mathbf{p}_e\mathbf{r}_1)(\mathbf{r}_1\mathbf{p}_e)}{r_1^4}
   \right]
   +\frac{Z_2^2\alpha^2}{8m^2M_2}
   \left[
      \mathbf{p}_e\frac{1}{r_2^2}\mathbf{p}_e+3\frac{(\mathbf{p}_e\mathbf{r}_2)(\mathbf{r}_2\mathbf{p}_e)}{r_2^4}
   \right],
\\[4mm]\displaystyle
\mathcal{U}_{6b} =
   -\frac{Z_1^2\alpha^2}{4m^2M_1}\>
      \frac{1}{r_1^4}\left[\mathbf{r}_1\!\times\!\mathbf{p}_e\right]\!\cdot\!\boldsymbol{\sigma}_e
   -\frac{Z_2^2\alpha^2}{4m^2M_2}\>
      \frac{1}{r_2^4}\left[\mathbf{r}_2\!\times\!\mathbf{p}_e\right]\!\cdot\!\boldsymbol{\sigma}_e\>,
\\[4mm]\displaystyle
\mathcal{U}_{6c} =
   \frac{Z_1^2\alpha^2}{4m^2M_1}\>\frac{1}{r_1^4}
   +\frac{Z_2^2\alpha^2}{4m^2M_2}\>\frac{1}{r_2^4}\>.
\end{array}
\end{equation}

\section{Effective Hamiltonian} \label{heff6}

In this Section, we collect the results obtained in Secs.~\ref{tree}-\ref{seagull} to build the complete effective Hamiltonian at orders $m\alpha^6$ and $m\alpha^6(m/M)$, including as well the second-order terms contributing to these orders. In doing so, we separate the different types of interactions: spin-independent, spin-orbit, spin-spin scalar and tensor interactions. Before that, we recall the expression of the effective Hamiltonian for leading-order relativistic corrections, i.e. the Breit-Pauli Hamiltonian, which comes into play in the second-order terms.

Spin-orbit interactions require a specific discussion. Formally, the leading electronic spin-orbit interaction $H_{so}$ (see Eq.~(\ref{breit-pauli}) below) contains terms of order $m\alpha^4$ (electronic spin-orbit) and $m\alpha^4(m/M)$ (electronic spin-nuclear orbit). However, assuming one considers a $\sigma$ electronic state, the electronic spin-orbit coupling gives a zero contribution in the Born-Oppenheimer approach. The nonzero value of this term is due to nonadiabatic effects, so that it is actually smaller by a factor $\sim (m/M)$ with respect to its nominal order, and thus of the same order as the electronic spin-nuclear orbit coupling. The same thing occurs in the relativistic corrections, i.e. spin-orbit terms that are nominally of order $m\alpha^6$ are of comparable magnitude to the ``recoil'' ($m\alpha^6(m/M)$ terms. We will thus make no distinction between nonrecoil or recoil contributions whenever the spin-orbit interaction is involved.

For the same reasons, the nuclear spin-orbit interaction $H_{so_N}$ (last line in Eq.~(\ref{breit-pauli})), in which the first term is of nominal order $m\alpha^4(m/M)$, has an actual contribution of order $m\alpha^4(m/M)^2$. Relativistic corrections to this interaction (e.g. the effective potential $\mathcal{U}_{2c}$, Eq.~(\ref{tr-ph})), are of order $m\alpha^6(m/M)^2$. That is why we will not consider the nuclear spin-orbit interaction in the following.

\subsection{Leading-order ($m\alpha^4$) relativistic corrections}

We include here all terms of the Breit-Pauli Hamiltonian at orders $m\alpha^4$ and $m\alpha^4(m/M)$ (the electron's anomalous magnetic moment is not taken into account here).
\begin{equation}
H^{(4)} = H_B + H_{ret} + H_{so} + H_{ss} + H_{so_N},
\end{equation}
\begin{equation} \label{breit-pauli}
\begin{array}{@{}l}\displaystyle
H_B = -\frac{p_e^4}{8m^3}+\frac{[\Delta_e V]}{8m^2},
\\[3mm]\displaystyle
H_{ret} =
   \frac{Z_1}{2}\>
      \frac{p_e^i}{m}\left(\frac{\delta^{ij}}{r_1}+\frac{r_1^ir_1^j}{r_1^3}\right)\frac{P_1^j}{M_1}
   +\frac{Z_2}{2}\>
      \frac{p_e^i}{m}\left(\frac{\delta^{ij}}{r_2}+\frac{r_2^ir_2^j}{r_2^3}\right)\frac{P_2^j}{M_2},
\\[3mm]\displaystyle
H_{so} =
   \frac{Z_a(1\!+\!2a_e)}{2m^2}\,\frac{[\mathbf{r}_a\!\times\!\mathbf{p}_e]}{r_a^3}\,\mathbf{s}_e
   -\frac{Z_a(1\!+\!a_e)}{mM_a}\,\frac{[\mathbf{r}_a\!\times\!\mathbf{P}_a]}{r_a^3}\,\mathbf{s}_e,
\\[3mm]\displaystyle
H_{ss} =
   \left[
      \frac{\boldsymbol{\mu}_e\boldsymbol{\mu}_a}{r_a^3}
      -3\frac{(\boldsymbol{\mu}_e\mathbf{r}_a)(\boldsymbol{\mu}_a\mathbf{r}_a)}{r_a^5}
   \right]
   -\frac{8\pi\alpha}{3}\boldsymbol{\mu}_e\boldsymbol{\mu}_a\delta(\mathbf{r}_a),
\\[3mm]\displaystyle
H_{so_N} =
   \frac{1}{m}\,\frac{[\mathbf{r}_a\!\times\!\mathbf{p}_e]}{r_a^3}\,\boldsymbol{\mu}_a
   -\frac{1}{M_a}\left[1-\frac{Zm_pI_a}{M_a\mu_a}\right] \, \frac{[\mathbf{r}_a\!\times\!\mathbf{P}_a]}{r_a^3}\,\boldsymbol{\mu}_a.
\end{array}
\end{equation}

\subsection{Spin-independent interaction: leading term and recoil}

The nonrecoil effective Hamiltonian is
\begin{equation}\label{nrec}
\begin{array}{@{}l}
\displaystyle
H_{no-spin}^{(6)} = \frac{p_e^6}{16m^5} + \mathcal{U}_{1a} + \mathcal{U}_4
\\[3mm] \displaystyle \hspace{8mm} = \frac{p_e^6}{16m^5}
     -\frac{3}{64m^4}\left\{p_e^2,\Delta V\right\}
     +\frac{5}{128m^4}\left\{p_e^4,V\right\}
     -\frac{5}{64m^4}\left(p_e^2Vp_e^2\right)
     + \frac{1}{8m^3}\mathbf{E}^2
     .
\end{array}
\end{equation}

The second-order contribution is
\begin{equation}
\Delta E_{no-spin}^{2^{nd}-order} =
\left\langle
   H_B\,Q (E_0-H_0)^{-1} Q\,H_B
\right\rangle.
\end{equation}

The above expressions coincide with previous results~\cite{Pachucki97,Korobov07}. These corrections have been evaluated numerically for the hydrogen molecular ions in~\cite{Korobov08,Korobov17}. It should be noted that both the first-order and second-order contributions contain divergences, which need to be cancelled out~\cite{Korobov07}.

The recoil effective Hamiltonian is
\begin{equation} \label{hrec}
\begin{array}{@{}l}\displaystyle
H_{rec}^{(6)} = \mathcal{U}_{2a} + \mathcal{U}_{3a} + \mathcal{U}_{6a} + \mathcal{U}_{6c}\>,
\\[3mm]\displaystyle
\mathcal{U}_{2a} =
   -\frac{Z_1}{8m^2}\>
      \left\{p_e^2,\frac{p_e^i}{m}\left(\frac{\delta^{ij}}{r_1}+\frac{r_1^ir_1^j}{r_1^3}\right)\frac{P_1^j}{M_1}\right\}
   -\frac{Z_2}{8m^2}\>
      \left\{p_e^2,\frac{p_e^i}{m}\left(\frac{\delta^{ij}}{r_2}+\frac{r_2^ir_2^j}{r_2^3}\right)\frac{P_2^j}{M_2}\right\},
\\[3mm]\displaystyle
\mathcal{U}_{3a} =
   \frac{1}{4m}
      \left[
         \frac{Z_1^3}{M_1}\frac{1}{r_1^3}+\frac{Z_2^3}{M_2}\frac{1}{r_2^3}
         +\frac{Z_1^2Z_2}{M_1}\frac{(\mathbf{r}_1\mathbf{r}_2)}{r_1^2r_2^3}
         +\frac{Z_1Z_2^2}{M_2}\frac{(\mathbf{r}_1\mathbf{r}_2)}{r_1^3r_2^2}
      \right]
\\[3mm]\hspace{8mm}\displaystyle
      +\frac{Z_1^2}{8m^2M_1}
         \left[
            \frac{1}{r_1^4}+\mathbf{p}_e\frac{1}{r_1^2}\mathbf{p}_e
               -3\frac{(\mathbf{p}_e\mathbf{r}_1)(\mathbf{r}_1\mathbf{p}_e)}{r_1^4}
         \right]
      +\frac{Z_2^2}{8m^2M_2}
         \left[
            \frac{1}{r_2^4}+\mathbf{p}_e\frac{1}{r_2^2}\mathbf{p}_e
               -3\frac{(\mathbf{p}_e\mathbf{r}_2)(\mathbf{r}_2\mathbf{p}_e)}{r_2^4}
         \right],
\\[3mm]\displaystyle
\mathcal{U}_{6a} =
   \frac{Z_1^2}{8m^2M_1}
   \left[
      \mathbf{p}_e\frac{1}{r_1^2}\mathbf{p}_e+3\frac{(\mathbf{p}_e\mathbf{r}_1)(\mathbf{r}_1\mathbf{p}_e)}{r_2^4}
   \right]
   +\frac{Z_2^2}{8m^2M_2}
   \left[
      \mathbf{p}_e\frac{1}{r_2^2}\mathbf{p}_e+3\frac{(\mathbf{p}_e\mathbf{r}_2)(\mathbf{r}_2\mathbf{p}_e)}{r_2^4}
   \right],
\\[3mm]\displaystyle
\mathcal{U}_{6c} =
   \frac{Z_1^2}{4m^2M_1}\>\frac{1}{r_1^4}
   +\frac{Z_2^2}{4m^2M_2}\>\frac{1}{r_2^4}\>.
\end{array}
\end{equation}

The second-order contribution is
\begin{equation}
\begin{array}{@{}l}\displaystyle
\Delta E_{rec}^{2^{nd}-order} = \Delta E_{ret} + \Delta E_{so\hbox{-}so}^{(0)},
\\[3mm]\displaystyle
\Delta E_{ret} =
2\left\langle
   H_B\,Q (E_0-H_0)^{-1} Q\,H_{ret}
\right\rangle
\\[3mm]\displaystyle
\Delta E_{so\hbox{-}so}^{(0)} =
   \left\langle
      H_{so} Q (E_0-H_0)^{-1} Q H_{so}
   \right\rangle^{(0)}.
\end{array}
\end{equation}
where $A^{(0)}$ denotes the scalar part of an operator $A$ ($H_{so}$ being a vector operator, the second-order term has contributions of rank 0, 1, and 2), see Appendix~\ref{2nd-order-algebra}, Eq.~(\ref{DeltaE2}) for details.

It should be noted that $\Delta E_{so\hbox{-}so}^{(0)}$ was considered together with nonrecoil terms in Refs.~\cite{Korobov07,Korobov08}. For the reasons explained above, we prefer to include it in the recoil part.

The effective Hamiltonian~(\ref{hrec}) is actually incomplete, because it does not include contributions from the contact terms of the NRQED Lagrangian~\cite{Caswell86,Kinoshita96,Patkos16}. A complete consideration of the recoil effective Hamiltonian for the hydrogen molecular ions, including contact terms and explicit cancellation of divergences, can be found in~\cite{Zhong18}. Our results coincide with those of that reference: the potentials $\mathcal{U}_{2a}$, $\mathcal{U}_{3a}$, and $\mathcal{U}_{6a}$ respectively appear in the terms denoted $\delta H_4$ (Eq.~(42)), $\delta H_6$ (Eq.~(50)), and $\delta H_5$ (Eq.~(45)). In the case of $\mathcal{U}_{3a}$, this is best seen by comparing the first line of Eq.~(\ref{U3a}) with Eq.~(50) of~\cite{Zhong18}, where a prefactor $z_a z_e$ should be added.

\subsection{Spin-spin scalar interaction}

The effective Hamiltonian for this interaction is
\begin{equation}
\begin{array}{@{}l} \displaystyle
H_{ss(6)}^{(0)} = \mathcal{U}_{2d}^{(0)} + \mathcal{U}_{5b}^{(0)},
\\[3mm]\displaystyle
\mathcal{U}_{2d}^{(0)} =
   \phantom{-}\frac{1}{4m^2}\,
      \left\{p_e^2,\left[\frac{8\pi}{3}\,\delta^3(\mathbf{r}_a)\right]\right\}
      \boldsymbol{\mu}_e\boldsymbol{\mu}_a
\\[3mm]\displaystyle
\mathcal{U}_{5b}^{(0)} =
    -\frac{1}{3m}
    \left[
       Z_1\frac{\boldsymbol{\mu}_e\boldsymbol{\mu}_1}{r_1^4}
       +Z_2\frac{\boldsymbol{\mu}_e\boldsymbol{\mu}_2}{r_2^4}
       +\frac{Z_2(\mathbf{r}_1\mathbf{r}_2)\boldsymbol{\mu}_e\boldsymbol{\mu}_1}{r_1^3r_2^3}
       +\frac{Z_1(\mathbf{r}_1\mathbf{r}_2)\boldsymbol{\mu}_e\boldsymbol{\mu}_2)}{r_1^3r_2^3}
    \right]
\end{array}
\end{equation}
and the second-order contribution is
\begin{equation}
\Delta E_{ss}^{(0)2^{nd}-order} =
2 \left\langle
   H_B\,Q (E_0-H_0)^{-1} Q\,H_{ss}^{(0)}
\right\rangle.
\end{equation}
Again, the first-order and second-order terms contain divergences which have to be cancelled out. This was done in~\cite{Korobov09,Korobov16} for hydrogen molecular ions, and the resulting corrections to the spin-spin contact interaction were evaluated numerically. Beyond the relativistic corrections considered here (of order $(Z\alpha)^2 E_F$, where $E_F$ is the Fermi splitting), there is a one-loop radiative contribution at the same order ($\alpha(Z\alpha)E_F$)~\cite{SapYen,EGS}. Other contributions to spin-spin scalar interactions include higher-order QED corrections (see~\cite{SapYen,EGS} and references therein) and effects involving the nuclear structure such as the Zemach~\cite{Zemach} and recoil~\cite{Arnowitt,BodYen88} corrections, and were taken into account in~\cite{Korobov09,Korobov16}.

\subsection{Electron spin-orbit interaction} \label{sec-so}

As explained above, for this interaction we make no distinction between nonrecoil and recoil contributions. The effective Hamiltonian is
\begin{equation} \label{heff-so}
\begin{array}{@{}l}\displaystyle
H_{so}^{(6)} = \mathcal{U}_{1b} + \mathcal{U}_{2b} + \mathcal{U}_{5a} + \mathcal{U}_{6b},
\\[3mm]\displaystyle
\mathcal{U}_{1b} =
   -\frac{3Z_1}{16m^4}
         \left\{p_e^2,\frac{1}{r_1^3}[\mathbf{r}_1\!\times\!\mathbf{p}_e]\right\}\mathbf{s}_e
   -\frac{3Z_2}{16m^4}
         \left\{p_e^2,\frac{1}{r_2^3}[\mathbf{r}_2\!\times\!\mathbf{p}_e]\right\}\mathbf{s}_e\,,
\\[3mm]\displaystyle
\mathcal{U}_{2b} =
   \frac{Z_1}{4m^3M_1}
   \left\{p_e^2,\frac{1}{r_1^3}\bigl[\mathbf{r}_1\!\times\!\mathbf{P}_1\bigr]\right\}\mathbf{s}_e
   +\frac{Z_2}{4m^3M_2}
   \left\{p_e^2,\frac{1}{r_2^3}\bigl[\mathbf{r}_2\!\times\!\mathbf{P}_2\bigr]\right\}\mathbf{s}_e\,,
\\[3mm]\displaystyle
\mathcal{U}_{5a} =
   \frac{Z_1^2}{4m^2M_1}\>
     \frac{1}{r_1^4}\left[\mathbf{r}_1\!\times\!\mathbf{P}_1\right]\mathbf{s}_e
   +\frac{Z_2^2}{4m^2M_2}\>
     \frac{1}{r_2^4}\left[\mathbf{r}_2\!\times\!\mathbf{P}_2\right]\mathbf{s}_e
\\[3mm]\displaystyle\hspace{10mm}
   +\frac{Z_1Z_2}{4m^2M_1}\>
     \frac{1}{r_1r_2^3}\left[\mathbf{r}_2\!\times\!\mathbf{P}_1\right]\mathbf{s}_e
   +\frac{Z_1Z_2}{4m^2M_2}\>
     \frac{1}{r_1^3r_2}\left[\mathbf{r}_1\!\times\!\mathbf{P}_2\right]\mathbf{s}_e
\\[3mm]\displaystyle\hspace{10mm}
   -\frac{Z_1Z_2}{4m^2M_1}\>
     \frac{1}{r_1^3r_2^3}\left[\mathbf{r}_1\!\times\!\mathbf{r}_2\right](\mathbf{r}_1\mathbf{P}_1)\mathbf{s}_e
   +\frac{Z_1Z_2}{4m^2M_2}\>
     \frac{1}{r_1^3r_2^3}\left[\mathbf{r}_1\!\times\!\mathbf{r}_2\right](\mathbf{r}_2\mathbf{P}_2)\mathbf{s}_e\,,
\\[3mm]\displaystyle
\mathcal{U}_{6b} =
   -\frac{Z_1^2}{2m^2M_1}\>
      \frac{1}{r_1^4}\left[\mathbf{r}_1\!\times\!\mathbf{p}_e\right]\mathbf{s}_e
   -\frac{Z_2^2}{2m^2M_2}\>
      \frac{1}{r_2^4}\left[\mathbf{r}_2\!\times\!\mathbf{p}_e\right]\mathbf{s}_e\,.
\end{array}
\end{equation}

The second-order contributions is
\begin{equation} \label{2nd-order-so}
\begin{array}{@{}l}\displaystyle
\Delta E_{so}^{2^{nd}-order} = \Delta E_{so} + \Delta E_{so\hbox{-}ret} + \Delta E_{so\hbox{-}so}^{(1)},
\\[3mm]\displaystyle
\Delta E_{so} =
   2\left\langle
      H_{so} Q (E_0-H_0)^{-1} Q H_B
   \right\rangle,
\\[3mm]\displaystyle
\Delta E_{so\hbox{-}ret} =
   2\left\langle
      H_{so} Q (E_0-H_0)^{-1} Q H_{ret}
   \right\rangle,
\\[3mm]\displaystyle
\Delta E_{so\hbox{-}so}^{(1)} =
   \left\langle
      H_{so} Q (E_0-H_0)^{-1} Q H_{so}
   \right\rangle^{(1)}.
\end{array}
\end{equation}
It is worth noting that both first and second-order terms are finite and do not require regularization. The spin-orbit interaction at this order was partially considered in~\cite{Korobov09b} for the antiprotonic helium atom, but all terms were not included in that work. It should also be mentioned that the only other correction to the spin-orbit interaction at the $m\alpha^6$ order corresponds to the effect of the electron's anomalous magnetic moment~\cite{EGS} and was therefore already included in~\cite{Bakalov06}.

\subsection{Spin-spin tensor interaction} \label{sec-tensor}

The effective Hamiltonian for spin-spin tensor interaction is
\begin{equation}
\begin{array}{@{}l}\displaystyle
H_{ss(6)}^{(2)} = \mathcal{U}_{2d}^{(2)} + \mathcal{U}_{5b}^{(2)},
\\[3mm]\displaystyle
\mathcal{U}_{2d}^{(2)} =
   \phantom{-}\frac{1}{4m^2}\,
      \left\{p_e^2,
      \left[
            -\frac{r_a^2\boldsymbol{\mu}_e\boldsymbol{\mu}_a
                 \!-\!3(\boldsymbol{\mu}_e\mathbf{r}_a)
                       (\boldsymbol{\mu}_a\mathbf{r}_a)}{r_a^5}
      \right]\right\}
\\[3mm]\displaystyle
\mathcal{U}_{5b}^{(2)} =
    -\frac{1}{6m}
    \left[
       Z_1\frac{r_1^2\boldsymbol{\mu}_e\boldsymbol{\mu}_1
             -3(\boldsymbol{\mu}_e\mathbf{r}_1)(\boldsymbol{\mu}_1\mathbf{r}_1)}{r_1^6}
       +Z_2\frac{r_2^2\boldsymbol{\mu}_e\boldsymbol{\mu}_2
             -3(\boldsymbol{\mu}_e\mathbf{r}_2)(\boldsymbol{\mu}_2\mathbf{r}_2)}{r_2^6}
    \right]
\\[3mm]\displaystyle\hspace{10.5mm}
    -\frac{1}{6m}\,
    \left[
       Z_2\frac{(\mathbf{r}_1\mathbf{r}_2)\boldsymbol{\mu}_e\boldsymbol{\mu}_1
             -3(\boldsymbol{\mu}_e\mathbf{r}_1)(\boldsymbol{\mu}_1\mathbf{r}_2)}{r_1^3r_2^3}
       +Z_1\frac{(\mathbf{r}_1\mathbf{r}_2)\boldsymbol{\mu}_e\boldsymbol{\mu}_2
             -3(\boldsymbol{\mu}_e\mathbf{r}_2)(\boldsymbol{\mu}_2\mathbf{r}_1)}{r_1^3r_2^3}
    \right].
\end{array}
\end{equation}

The second-order contribution is
\begin{equation}
\begin{array}{@{}l}\displaystyle
\Delta E_{ss}^{(2)2^{nd}-order} = \Delta E_{ss}^{(2)} + \Delta E_{so\hbox{-}ss} + \Delta E_{so\hbox{-}so_N}^{(2)},
\\[3mm]\displaystyle
\Delta E_{ss}^{(2)} =
   2\left\langle
      H_{ss}^{(2)} Q (E_0-H_0)^{-1} Q H_B
   \right\rangle,
\\[3mm]\displaystyle
\Delta E_{so\hbox{-}ss}^{(2)} =
   2\left\langle
      H_{ss}^{(2)} Q (E_0-H_0)^{-1} Q H_{so}
   \right\rangle^{(2)},
\\[3mm]\displaystyle
\Delta E_{so\hbox{-}so_N}^{(2)} =
   2\left\langle
      H_{so} Q (E_0-H_0)^{-1} Q H_{so_N}
   \right\rangle^{(2)}.
\end{array}
\end{equation}
Here also, all terms are finite and do not require regularization. The only other correction to the spin-spin tensor interaction at this order is the effect of the electron's anomalous magnetic moment.

\section{Numerical results and conclusion} \label{results}

The results of Secs.~\ref{sec-so} and~\ref{sec-tensor} can be used to calculate relativistic corrections to the electron spin-orbit and spin-spin tensor interaction coefficients of the hyperfine Hamiltonian in a one-electron molecular system. Here, we present corrections to the spin-orbit coefficient in both H$_2^+$~\cite{Korobov06} and HD$^+$~\cite{Bakalov06} for a few transitions studied in recent or ongoing experiments~\cite{Alighanbari20,Patra19,Louvradoux19,Zhong15}.

Our calculations rely on the ``exponential'' variational expansion~\cite{Korobov00}, where the wave function for a state of total orbital angular momentum $L$ and parity $\Pi=(-1)^L$ is expanded in the following way:
\begin{equation}\label{var_expansion}
\begin{array}{@{}l}
\displaystyle \Psi_{LM}^\pi(\mathbf{R},\mathbf{r}_1) =
       \sum_{l_1+l_2=L}
         \mathcal{Y}^{l_1l_2}_{LM}(\hat{\mathbf{R}},\hat{\mathbf{r}}_1)
         G^{L\pi}_{l_1l_2}(R,r_1,r_2),
\\[5mm]\displaystyle
\mathcal{Y}^{l_1l_2}_{LM}(\hat{\mathbf{R}},\hat{\mathbf{r}}_1) = R^{l_1} r_1^{l_2} \left\{ Y_{l_1}(\hat{\mathbf{R}}) \otimes Y_{l_2}(\hat{\mathbf{r}}_1) \right\}_{LM},
\\[1mm]\displaystyle
G_{l_1l_2}^{L\pi}(R,r_1,r_2) = \sum_{n=1}^N \Big\{C_n\,\mbox{Re} \bigl[e^{-\alpha_n R-\beta_n r_1-\gamma_n r_2}\bigr]
+D_n\,\mbox{Im} \bigl[e^{-\alpha_n R-\beta_n r_1-\gamma_n r_2}\bigr] \Big\}.
\end{array}
\end{equation}
The complex exponents $\alpha_n$, $\beta_n$, $\gamma_n$ are generated in a pseudorandom way. Matrix elements of all operators are calculated analytically; general methods of such calculations may be found e.g. in~\cite{Drake06,Drake78,Harris04}. The matrix elements are reduced to finite sums of radial integrals of the general form
\begin{equation}
\Gamma_{l,m,n}(\alpha,\beta,\gamma) = \int_0^{\infty} dR \int_0^{\infty} dr_1 \int_{|R - r_1|}^{R + r_1} dr_{2} R^l r_1^m r_2^n \, e^{-\alpha R - \beta r_1 - \gamma r_2},
\end{equation}
which are then calculated using recurrence relations.

From the numerical point of view, the calculation of first-order terms ($\mathcal{U}_{1b}$, $\mathcal{U}_{2b}$, and $\mathcal{U}_{5a}$) is straightforward; we used $N = 2000-3000$ which was more than sufficient to get 4 significant digits.

Second-order contributions pose more difficult problems~\cite{Korobov09b}, especially the singular term $\Delta E_{so}$ and the (less singular) term $\Delta E_{so-so}^{(1)}$. In the case of $\Delta E_{so}$, the intermediate wavefunction $\psi^{(1)}$ defined by
\begin{equation} \label{psi1}
(E_0 - H_0) \psi^{(1)} = (H_B - \langle H_B \rangle) \, \psi_0
\end{equation}
behaves like $1/r_1$ ($1/r_2$) at small electron-nucleus distances. The regular trial functions~~(\ref{var_expansion}) would thus result in very slow convergence. In order to reduce this singularity, we use the transformation described in~\cite{Korobov09b}:
\begin{equation}
H'_B = H_B - (E_0 - H_0) U - U(E_0 - H_0),
\end{equation}
where
\begin{equation}
U = \frac{c_1}{r_1} + \frac{c_2}{r_2}\,, \qquad c_i = \frac{\mu_i (2 \mu_i - m_e)}{4 m_e^3} Z_i.
\end{equation}
Here, $1/\mu_i = 1/m_e + 1/M_i$. The second-order term may then be rewritten as follows:
\begin{equation}
\left\langle H_{so} Q (E_0-H_0)^{-1} Q H_B \right\rangle = \left\langle H_{so} Q (E_0-H_0)^{-1} Q H'_B \right\rangle + \langle U H_{so} \rangle - \langle U \rangle \langle H_{so} \rangle.
\end{equation}
With $H_B$ being replaced by $H_B'$ in Eq.~(\ref{psi1}), the first-order wavefunction is now less singular and behaves like $\ln(r_1)$ ($\ln(r_2)$) at small distances. In the numerical evaluation, we use a ``multilayer'' basis set, where the first subsets (between 2 and 4) approximate the regular part of the intermediate wavefunction, and 8 others subsets contain growing exponents $\beta_n$ ($\gamma_n$) up to 10$^4$ in order to reproduce the $\ln(r_1)$ ($\ln(r_2)$) behavior at small distances, see an illustrative example in Table~\ref{param}. The total size of the intermediate basis set is typically around $N \sim 10000$. The convergence will be analyzed in more detail in a future publication focusing on numerical results.

\begin{table} [h!]
\small
\begin{tabular}{@{\hspace{1mm}}c@{\hspace{2mm}}@{\hspace{2mm}}c@{\hspace{2mm}}@{\hspace{2mm}}c@{\hspace{2mm}}@{\hspace{2mm}}c@{\hspace{2mm}}@{\hspace{2mm}}c@{\hspace{2mm}}@{\hspace{2mm}}c@{\hspace{2mm}}}
\hline\hline
Subset & $[A_1,A_2]$ & $[A'_1,A'_2]$ & $[B_1,B_2]$ & $[C_1,C_2]$ & $N_i$ \\
 \hline
 1 & $[5.1,5.5]$ & $[0.8,15.2]$  & $[0.00,1.80]$ & $[0.00,1.65]$ & $1830$ \\
 2 & $[5.1,5.2]$ & $[-0.6,6.3]$  & $[0.00,1.56]$ & $[0.00,1.59]$ & $1170$ \\
 3 & $[5.1,5.5]$ & $[-0.6,15.2]$ & $[1.80,10.0]$ & $[0.00,1.65]$ & $1290$ \\
 4 & $[5.1,5.5]$ & $[-0.6,15.2]$ & $[0.00,1.80]$ & $[1.65,10.0]$ & $1290$ \\
 5 & $[5.1,5.5]$ & $[-0.6,15.2]$ & $[10.0,10^2]$ & $[0.00,1.65]$ & $1070$ \\
 6 & $[5.1,5.5]$ & $[-0.6,15.2]$ & $[0.00,1.80]$ & $[10.0,10^2]$ & $1070$ \\
 7 & $[5.1,5.5]$ & $[-0.6,15.2]$ & $[10^2,10^3]$ & $[0.00,1.65]$ & $900$  \\
 8 & $[5.1,5.5]$ & $[-0.6,15.2]$ & $[0.00,1.80]$ & $[10^2,10^3]$ & $900$  \\
 9 & $[5.1,5.5]$ & $[-0.6,15.2]$ & $[10^3,10^4]$ & $[0.00,1.65]$ & $740$  \\
10 & $[5.1,5.5]$ & $[-0.6,15.2]$ & $[0.00,1.80]$ & $[10^3,10^4]$ & $740$  \\
\hline\hline
\end{tabular}
\caption{\label{param} Variational parameters used in the calculation of singular second-order terms for the $(L=1,v=0)$ state of HD$^+$. $[A_1,A_2]$ (resp.~$[A'_1,A'_2]$ are the intervals in which the real (resp. imaginary) parts of exponents $\alpha_n$ (see Eq.~(\ref{var_expansion})) are generated, and $[B_1,B_2]$ (resp. $[C_1,C_2]$) are the intervals for the real parts of $\beta_n$ (resp. $\gamma_n$). An indicative number of basis functions $N_i$ in each subset is given the last column. The total basis size in this example is $N = 11000$.}
\end{table}

The $\Delta E_{so-so}^{(1)}$ contribution is obtained from Eq.~(\ref{DeltaE2}) of the Appendix. The spin operator $\mathbf{U}^1$ appearing in that equation is given in Sec.~\ref{spin-op}. For the orbital operator $\mathbf{T}^1$, the calculation is separated into three terms $a_0$, $a_-$, $a_+$ corresponding to the possible values of the angular momentum $L'$ of intermediate states, $L' = L,L \pm 1$, see Eq.~(\ref{components}) for definitions. The total contribution is given by Eq.~(\ref{mat-elem}) (with $k_1 = k_2 = k = 1$). Since the first-order wavefunction~(\ref{psi1}) with $H_{so}$ on the right-hand side is also singular at small electron-nucleus distances, we use a similar multilayer basis set as for the $\Delta E_{so}$ contribution. An intermediate basis size up to $N \sim 20000$ was used for the $(L=3,v=9)$ state of HD$^+$.

Our numerical results are given in Tables~\ref{table-h2plus} (for H$_2^+$) and~\ref{table-hdplus} (for HD$^+$). From a study of convergence as a function of $N$, we estimate the numerical uncertainty of $\Delta E_{so}$ and $\Delta E_{so-so}$ to about 1~Hz. For the other contributions, all digits are significant. The total uncertainty of the spin-orbit interaction coefficient (denoted $c_e$ in H$_2^+$ and $E_1$ in HD$^+$) is dominated by the yet unevaluated radiative correction of order $m \alpha^7 \ln(\alpha)$~\cite{Pachucki99,Jentschura05}. A tentative order of magnitude is $\alpha^3 \ln(\alpha) c_e \sim 100$~Hz, but our preliminary calculations indicate that this correction is actually as large as 300-400~Hz. The calculation of this contribution is thus essential for further improvement of theoretical predictions of the hyperfine structure, and will be addressed in a forthcoming publication.

\begin{table} [h!]
\small
\begin{tabular}{|@{\hspace{1mm}}c@{\hspace{1mm}}|@{\hspace{1mm}}c@{\hspace{1mm}}|@{\hspace{1mm}}c@{\hspace{1mm}}|@{\hspace{1mm}}c@{\hspace{1mm}}
|@{\hspace{1mm}}c@{\hspace{1mm}}|@{\hspace{1mm}}c@{\hspace{1mm}}|@{\hspace{1mm}}c@{\hspace{1mm}}|@{\hspace{1mm}}c@{\hspace{1mm}}
|@{\hspace{1mm}}c@{\hspace{1mm}}|@{\hspace{1mm}}c@{\hspace{1mm}}|c@{\hspace{1mm}}|}
\hline
$(L,v)$ & $c_e^{(BP)}$ & $U_{1b}$ & $U_{2b}$ & $U_{5a}$ & $\Delta E_{so}$ & $\Delta E_{so-ret}$ & $\Delta E_{so-so}^{(1)}$ & $\Delta c_e^{(6)}$ & $c_e$(this work)  \\
\hline
(2,0)   & $42\,162.530$ & 1.542 & $-$3.601 & 0.027 & 2.736 & 0.348 & 0.412 & {\bf 1.463} & $42\,163.99$ \\
\hline
(2,1)   & $39\,571.598$ & 1.451 & $-$3.440 & 0.036 & 2.579 & 0.327 & 0.388 & {\bf 1.341} & $39\,572.94$ \\
\hline
\end{tabular}
\caption{\label{table-h2plus} Relativistic corrections to the spin-orbit interaction coefficient $c_e$ for rovibrational states of H$_2^+$ (in kHz). The leading-order (Breit-Pauli) value of $c_e$ (Ref.~\cite{Korobov06}) is given in column 2. Columns 3-8 are the first-order and second-order contributions listed in Eqs.~(\ref{heff-so}) and~(\ref{2nd-order-so}), respectively. The total correction is given in column 9.  The last column is our new value of $c_e$.}
\end{table}

\begin{table} [h!]
\small
\begin{tabular}{|@{\hspace{1mm}}c@{\hspace{1mm}}|@{\hspace{1mm}}c@{\hspace{1mm}}|@{\hspace{1mm}}c@{\hspace{1mm}}|@{\hspace{1mm}}c@{\hspace{1mm}}
|@{\hspace{1mm}}c@{\hspace{1mm}}|@{\hspace{1mm}}c@{\hspace{1mm}}|@{\hspace{1mm}}c@{\hspace{1mm}}|@{\hspace{1mm}}c@{\hspace{1mm}}
|@{\hspace{1mm}}c@{\hspace{1mm}}|@{\hspace{1mm}}c@{\hspace{1mm}}|c@{\hspace{1mm}}|}
\hline
$(L,v)$ & $E_1^{(BP)}$ & $U_{1b}$ & $U_{2b}$ & $U_{5a}$ & $\Delta E_{so}$ & $\Delta E_{so-ret}$ & $\Delta E_{so-so}^{(1)}$ & $\Delta E_1^{(6)}$ & $E_1$(this work)  \\
\hline
(1,0)   & $31\,984.645$ & 1.170 & $-$2.736 & 0.021 & 2.087 & 0.263 & 0.313 & {\bf 1.118} & $31\,985.76$  \\
\hline
(1,6)   & $22\,643.474$ & 0.834 & $-$2.097 & 0.044 & 1.509 & 0.181 & 0.219 & {\bf 0.689} & $22\,644.16$  \\
\hline
(3,0)   & $31\,627.353$ & 1.156 & $-$2.694 & 0.019 & 2.043 & 0.260 & 0.308 & {\bf 1.093} & $31\,628.45$  \\
\hline
(3,9)   & $18\,270.577$ & 0.680 & $-$1.732 & 0.043 & 1.161 & 0.146 & 0.182 & {\bf 0.481} & $18\,271.06$  \\
\hline
\end{tabular}
\caption{\label{table-hdplus} Same as Table~\ref{table-h2plus}, for the spin-orbit coefficient  $E_1$ in HD$^+$. The Breit-Pauli value in column 2 was obtained in Ref.~\cite{Bakalov06}.}
\end{table}

In conclusion, we have derived the complete effective Hamiltonian at the $m\alpha^6$ and $m\alpha^6(m/M)$ for hydrogen molecular ions. The spin-independent and spin-spin scalar interaction terms were found to agree with previous calculations~\cite{Korobov09,Korobov08,Zhong18}. We then exploited this effective Hamiltonian to calculate corrections to the electronic spin-orbit hyperfine coefficient for a few states involved in experimentally studied transitions in H$_2^+$ and HD$^+$. The theoretical uncertainty has been reduced by more than a factor of 3, from about $\alpha^2 c_e \sim 1.5$~kHz to about 300-400~Hz. Next steps are the calculation of radiative corrections at the next order, and of corrections to the spin-spin tensor interaction coefficients. It will then become possible to perform precise comparison with present and upcoming experimental data. Finally, the effective Hamiltonian we have derived may also be used to improve the hyperfine structure calculations in antiprotonic helium~\cite{Korobov09b,Hori16}.

\section*{Acknowledgements}

The authors thank L. Hilico for useful comments on the manuscript. V.I.K. acknowledges the support of the Russian Foundation for Basic Research under Grant No. 19-02-00058-a. Z.-X.Z. acknowledges support from the National Natural Science Foundation of China (Grants No. 91636216, No. 11974382, and No. 11474316), and from the Chinese Academy of Sciences (the Strategic Priority Research Program, Grant No. XDB21020200, and the YIPA program).

\appendix

\section{Fourier integrals} \label{Fourier}

In this Appendix, we summarize the three-dimensional integrals that were used in our derivations for the Fourier transformation from momentum to coordinate space. The master integral is
\begin{equation}\label{ap:Coulomb}
\frac{4\pi}{(2\pi)^3}\int \frac{d^3\mathbf{q}}{\mathbf{q}^2}\>e^{i\mathbf{qr}} =
   \frac{1}{r}\,,
\end{equation}
and other useful integrals are
\begin{equation} \label{ap:Fourier}
\begin{array}{@{}l}\displaystyle
\frac{4\pi}{(2\pi)^3}\int\frac{d\mathbf{q}}{\mathbf{q}^2}
   \left(\delta^{ij}\!-\!\frac{q^iq^j}{\mathbf{q}^2}\right)\>e^{i\mathbf{qr}}  =
   \frac{1}{2}\left[\frac{\delta^{ij}}{r}+\frac{r^ir^j}{r^3}\right] ,
\\[3mm]\displaystyle
\frac{4\pi}{(2\pi)^3}\int\frac{d\mathbf{q}}{\mathbf{q}^4}
        \left(e^{i\mathbf{qr}}-1\right) = -\frac{r}{2}\,,
\\[3mm]\displaystyle
\frac{4\pi}{(2\pi)^3}\int\frac{d\mathbf{q}}{\mathbf{q}^4}
        \left(\delta^{ij}\!-\!\frac{q^iq^j}{\mathbf{q}^2}\right)
        \left(e^{i\mathbf{qr}}-1\right)
 = \frac{1}{8r}\left(r^ir^j-3r^2\delta^{ij}\right) ,
\\[3mm]\displaystyle
\frac{4\pi}{(2\pi)^3}\int\frac{d\mathbf{q}}{\mathbf{q}^2}\>
   [\mathbf{a}\!\times\!\mathbf{q}][\mathbf{b}\!\times\!\mathbf{q}]\>e^{i\mathbf{qr}} =
   -\left[\frac{(\mathbf{ab})}{r^3}-3\frac{(\mathbf{ar})(\mathbf{br})}{r^5}\right]
   +\frac{8\pi}{3}(\mathbf{ab})\,\delta(\mathbf{r})\,.
\end{array}
\end{equation}

\section{Algebra of angular momenta for the second-order contributions} \label{2nd-order-algebra}

A second-order contribution to the hyperfine splitting of a rovibrational state $(v,L)$ may be written in the general form
\begin{equation}
\Delta E =
   \left\langle
     vLSJM \left| \, (\mathbf{S}^{k_1}_a\cdot\mathbf{O}^{k_1}_a)Q(E_0-H_0)^{-1}Q(\mathbf{O}^{k_2}_b\cdot\mathbf{S}^{k_2}_b) \right| vLS'JM
   \right\rangle
\end{equation}
where $\mathbf{S}_a$, $\mathbf{S}_b$, $\mathbf{O}_a$, $\mathbf{O}_b$ are some irreducible orbital tensor operators, with $\mathbf{S}_a$, $\mathbf{S}_b$ acting in the spin space and $\mathbf{O}_a$, $\mathbf{O}_b$ in the orbital space. $|vLSJM\rangle$ is a pure hyperfine state, with $\mathbf{S}$ the total spin, and $\mathbf{J} = \mathbf{L} + \mathbf{S}$ the total angular momentum.

The goal of this Appendix is to show how such quantities can be decomposed into irreducible tensor components, which are expressed as the scalar product of an irreducible orbital tensor operator with an irreducible spin operator of the same rank. Then, we will give the expressions of the spin operators and of the orbital reduced matrix elements.

\subsection{Decomposition into irreducible tensor components}

Let us introduce the irreducible tensor operators
\[
\begin{array}{@{}l}\displaystyle
T^k_M = \{O^{k_1}_a\otimes Q (E_0-H_0)^{-1} Q O^{k_2}_b\}_{kM},
\qquad
U^k_M = \{S^{k_1}_a\otimes S^{k_2}_b\}_{kM}.
\end{array}
\]
Then, using the relationship (see Ref.~\cite{Varsh}, Chapter~3, Sec.~3.3.2, Eq.~(11))
\[
\{\{\mathbf{A}_{k_1}\otimes\mathbf{B}_{k_1}\}_0
   \otimes\{\mathbf{C}_{k_2}\otimes\mathbf{D}_{k_2}\}_0\}_{00}
= \sum_k \frac{\Pi_k}{\Pi_{k_1k_2}}
  \{\{\mathbf{A}_{k_1}\otimes\mathbf{C}_{k_2}\}_k
   \otimes\{\mathbf{B}_{k_1}\otimes\mathbf{D}_{k_2}\}_k\}_{00},
\]
where $\Pi_{n_1n_2\ldots} = \sqrt{(2n_1+1)(2n_2+1)\ldots}$, one gets
\begin{equation}
\begin{array}{@{}l}\displaystyle
(\mathbf{S}^{k_1}_a\cdot\mathbf{O}^{k_1}_a) Q (E_0-H_0)^{-1} Q (\mathbf{O}^{k_2}_b\cdot\mathbf{S}^{k_2}_b) =
\\[3mm]\displaystyle\hspace{15mm}
 = (-1)^{k_1 + k_2} \Pi_{k_1 k_2} \left\{ \{ \mathbf{S}^{k_1}_a \otimes \mathbf{O}^{k_1}_a \}_{0} \otimes \{ Q (E_0-H_0)^{-1} Q \mathbf{O}^{k_2}_b\cdot\mathbf{S}^{k_2}_b \}_{0} \right\}_{00}
\\[3mm]\displaystyle\hspace{15mm}
= (-1)^{k_1+k_2} \sum_k \Pi_k \{ \mathbf{T}^k \otimes \mathbf{U}^k \}_{00} = \sum_k (-1)^{k_1+k_2+k} (\mathbf{T}^k \cdot \mathbf{U}^k).
\end{array}
\end{equation}
As a result,
\begin{equation} \label{DeltaE2}
\begin{array}{@{}l}\displaystyle
\Delta E = \sum_k \Delta E^{(k)},
\\[3mm]\displaystyle
\Delta E^{(k)} =
   (-1)^{k_1 + k_2 + k} \left\langle vLSJM\left|(\mathbf{T}^k\cdot\mathbf{U}^k)\right|vLS'JM\right\rangle
\\[3mm]\displaystyle\hspace{11mm}
 = (-1)^{k_1 + k_2 + k}
   \frac{\left\langle vL\| \mathbf{T}^k \| vL\right\rangle}
        {\left\langle L\| \mathbf{L}^k \| L\right\rangle}
   \left\langle vLSJM\left|(\mathbf{L}^k\cdot\mathbf{U}^k)\right|vLS'JM\right\rangle\>,
\end{array}
\end{equation}
where $\mathbf{L}^0 = I$, $\mathbf{L}^1 = \mathbf{L}$, $\mathbf{L}^2 = \{ \mathbf{L} \otimes \mathbf{L} \}_{2\mu}$, etc.

\subsection{Irreducible spin operators} \label{spin-op}

\noindent a) With the electron spin-orbit Hamiltonian $H_{so}$ on both sides:
\[
\begin{array}{@{}l}\displaystyle
\mathbf{U}_0 = \{\mathbf{s}_e\otimes\mathbf{s}_e\}_{00} = -\frac{1}{\sqrt{3}}\,\mathbf{s}_e^2 = -\frac{\sqrt{3}}{4},
\\[3mm]\displaystyle
\mathbf{U}_1 = \{\mathbf{s}_e\otimes\mathbf{s}_e\}_{1\mu} = -\frac{1}{\sqrt{2}}\,\mathbf{s}_e,
\\[4mm]\displaystyle
\mathbf{U}_2 \equiv 0.
\end{array}
\]
b) With the electron spin-orbit Hamilotnian $H_{so}$ and the nuclear spin-orbit Hamiltonian $H_{so_N}$:
\[
\begin{array}{@{}l}\displaystyle
\mathbf{U}_0 = \{\mathbf{s}_e\otimes\mathbf{I}_a\}_{00} = -\frac{1}{\sqrt{3}}\,(\mathbf{s}_e\cdot\mathbf{I}_a),
\\[3mm]\displaystyle
\mathbf{U}_1 = \{\mathbf{s}_e\otimes\mathbf{I}_a\}_{1\mu} = \frac{i}{\sqrt{2}}\,[\mathbf{s}_e\times \mathbf{I}_a],
\\[3mm]\displaystyle
\mathbf{U}_2 = \{\mathbf{s}_e\otimes\mathbf{I}_a\}_{2\mu}
 = \sqrt{\frac{3}{2}}
   \left[\frac{1}{2}(s_e^iI_a^j+s_e^jI_a^j)-\frac{\delta^{ij}}{3}(\mathbf{s}_e\cdot\mathbf{I}_a)\right]_{2\mu}.
\end{array}
\]
c) With the electron spin-orbit Hamiltonian $H_{so}$ and the tensor spin-spin Hamiltonian $H_{ss}^{(2)}$:

\noindent Let us define:
\[
\mathbf{S}_{ss}^{(2)} = \{\mathbf{s}_e\otimes\mathbf{I}_a\}_{2\mu}.
\]
We then get
\[
\begin{array}{@{}l}\displaystyle
\mathbf{U}_1 = \{\mathbf{s}_e\otimes\mathbf{S}_{ss}\}_{1\mu} =
\{\mathbf{s}_e\otimes\{\mathbf{s}_e\otimes\mathbf{I}_a\}_2\}_{1\mu}
\\[3mm]\displaystyle \hspace{6mm}
= \frac{\sqrt{5}}{3}\{\{\mathbf{s}_e\otimes\mathbf{s}_e\}_0\otimes\mathbf{I}_a\}_{1\mu}
   +\frac{\sqrt{15}}{6}\{\{\mathbf{s}_e\otimes\mathbf{s}_e\}_1\otimes\mathbf{I}_a\}_{1\mu}
 = -\frac{\sqrt{15}}{12}\mathbf{I}_a - i\frac{\sqrt{15}}{12}[\mathbf{s}_e\!\times\!\mathbf{I}_a],
\\[3mm]\displaystyle
\mathbf{U}_2 = \{\mathbf{s}_e\otimes\mathbf{S}_{ss}\}_{2\mu} = -\frac{1}{2} \sqrt{\frac{3}{2}} \mathbf{S}_{ss}^{(2)},
\\[4mm]\displaystyle
\mathbf{U}_3 \equiv 0.
\end{array}
\]

\subsection{Orbital reduced matrix elements}

For the operator $T^k$ acting on spatial degrees of freedom, one separates the calculation into different terms corresponding to the possible vales of the angular momentum $L'$ of intermediate states, $L' = L,L \pm 1$ (since ${\rm min}(k_1,k_2) = 1$ in all the cases under consideration here). The reduced matrix element of a given component $L'$ may then be expressed (see Ref.~\cite{Varsh}, Chapter~13, Sec.~13.1.3, Eq.~(10))
\begin{equation}
\begin{array}{@{}l}\displaystyle
\left\langle vL\| \mathbf{T}^{k(L')} \| vL\right\rangle =
   (-1)^k\Pi_{k}
   \left\{\begin{matrix}
      k_1 & k_2 & k \\ L & L & L'
   \end{matrix}\right\}
   \sum_{n\ne0} \frac{\left\langle vL\|\mathbf{O}_a^{k_1}\|v_nL'\right\rangle
   \left\langle v_nL'\|\mathbf{O}_b^{k_2}\|vL\right\rangle}{E_0-E_n}\,.
\end{array}
\end{equation}
Let us define
\begin{equation} \label{components}
\begin{array}{@{}l}\displaystyle
a_- = -\frac{1}{\Pi_L^2}
   \sum_{n\ne0} \frac{\left\langle vL\|\mathbf{O}_a^{k_1}\|v_nL-1\right\rangle
   \left\langle v_nL-1\|\mathbf{O}_b^{k_2}\|vL\right\rangle}{E_0-E_n}\,,
\\[3mm]\displaystyle
a_0 = \frac{1}{\Pi_L^2}
   \sum_{n\ne0} \frac{\left\langle vL\|\mathbf{O}_a^{k_1}\|v_nL\right\rangle
   \left\langle v_nL\|\mathbf{O}_b^{k_2}\|vL\right\rangle}{E_0-E_n}\,,
\\[3mm]\displaystyle
a_+ = -\frac{1}{\Pi_L^2}
   \sum_{n\ne0} \frac{\left\langle vL\|\mathbf{O}_a^{k_1}\|v_nL+1\right\rangle
   \left\langle v_nL+1\|\mathbf{O}_b^{k_2}\|vL\right\rangle}{E_0-E_n}\,.
\end{array}
\end{equation}
The prefactor of the spin operator $\mathbf{L}^k\cdot\mathbf{U}^k$ in Eq.~(\ref{DeltaE2}) is given by
\begin{equation} \label{mat-elem}
\begin{array}{@{}l}\displaystyle
(-1)^{k_1 + k_2 + k}
   \frac{\left\langle vL\| \mathbf{T}^k \| vL\right\rangle}
        {\left\langle L\| \mathbf{L}^k \| L\right\rangle} =
\\[3mm]\displaystyle\hspace{15mm}
   (-1)^{k_1+k_2} \frac{\Pi_L^2\Pi_k}{\left\langle L\| \mathbf{L}^k \| L\right\rangle} \left[ -\left\{\begin{matrix}
      k_1 & k_2 & k \\ L & L & L-1
   \end{matrix}\right\} a_- + \left\{\begin{matrix}
      k_1 & k_2 & k \\ L & L & L
   \end{matrix}\right\} a_0 - \left\{\begin{matrix}
      k_1 & k_2 & k \\ L & L & L+1
   \end{matrix}\right\} a_+ \right].
\end{array}
\end{equation}


\begin{thebibliography}{99}

\bibitem{Karr16} J.-Ph.~Karr, L.~Hilico, J.C.J.~Koelemeij, and V.I.~Korobov,
   Hydrogen molecular ions for improved determination of fundamental constants.
   Phys. Rev. A \textbf{94}, 050501(R) (2016).
\bibitem{Alighanbari20} S.~Alighanbari, G.S.~Giri, F.L.~Constantin, V.I.~Korobov, and S.~Schiller,
   Precise test of quantum electrodynamics and determination of fundamental constants with HD$^+$ ions.
   Nature \textbf{581}, 152 (2020).
\bibitem{Tran13} V.Q.~Tran, J.-Ph.~Karr, A.~Douillet, J.C.J.~Koelemeij, and L.~Hilico,
   Two-photon spectroscopy of trapped HD$^+$ ions in the Lamb-Dicke regime.
   Phys. Rev. A~{\bf 88}, 033421 (2013).
\bibitem{Patra19} S.~Patra,
   Towards Doppler-free two-photon spectroscopy of trapped and cooled HD$^+$ ions.
   PhD thesis (VU Amsterdam, 2019).
\bibitem{Karr08} J.-Ph.~Karr, F.~Bielsa, A.~Douillet, J.~Pedregosa Gutierrez, V.I.~Korobov, and L.~Hilico,
   Vibrational spectroscopy of H$_2^+$: Hyperfine structure of two-photon transitions.
   Phys. Rev. A \textbf{77}, 063410 (2008).
\bibitem{Louvradoux19} T.~Louvradoux, PhD thesis (Universit\'e Paris Diderot, 2019).
\bibitem{Bakalov06} D. Bakalov, V.I. Korobov, and S. Schiller,
   High-Precision Calculation of the Hyperfine Structure of the HD$^+$ Ion.
   Phys. Rev. Lett.~\textbf{97}, 243001 (2006).
\bibitem{Korobov06} V.I. Korobov, L. Hilico, and J.-Ph. Karr,
   Hyperfine structure in the hydrogen molecular ion.
   Phys. Rev. A~\textbf{74}, 040502(R) (2006).
\bibitem{Korobov09} V.I. Korobov, L. Hilico, and J.-Ph. Karr,
   Relativistic corrections of $m\alpha^6(m/M)$ order to the hyperfine structure of the H$_2^+$ molecular ion.
   Phys. Rev. A~\textbf{79}, 012501 (2009).
\bibitem{Korobov16} V.I. Korobov, J.C.J. Koelemeij, L. Hilico, and J.-Ph. Karr,
   Theoretical Hyperfine Structure of the Molecular Hydrogen Ion at the 1 ppm Level.
   Phys. Rev. Lett.~\textbf{116}, 053003 (2016).
\bibitem{Jefferts69} K.B.~Jefferts,
   Hyperfine structure in the molecular ion H$_2^+$.
   Phys. Rev. Lett.~{\bf 23}, 1476 (1969).
\bibitem{Caswell86} W.E.~Caswell and G.P.~Lepage,
   Effective Lagrangians for bound state problems in QED, QCD, and othe field theories.
   Phys.\ Lett. B~\textbf{167}, 437 (1986).
\bibitem{Kinoshita96} T.~Kinoshita and M.~Nio,
   Radiative corrections to the muonium hyperfine structure: The $\alpha^2(Z\alpha)$ correction.
   Phys.\ Rev.~D \textbf{53}, 4909 (1996).
\bibitem{Hill13} R.J.~Hill, G.~Lee, G.~Paz, and M.P.~Solon,
   NRQED Lagrangian at order $1/M^4$.
   Phys.\ Rev. D~\textbf{87}, 053017 (2013).
\bibitem{Haidar20} M.~Haidar, Z.-X.~Zhong, V.I.~Korobov, and J.-Ph.~Karr,
   Nonrelativistic QED approach to the fine- and hyperfine-structure corrections of order $m\alpha^6$ and $m\alpha^6(m/M)$: Application to the hydrogen atom.
   Phys. Rev. A~\textbf{101}, 022501 (2020).
\bibitem{Patkos16} V.~Patk\'o\v{s}, V.A.~Yerokhin, and K.~Pachucki,
   Higher-order recoil corrections for triplet states of the helium atom.
   Phys.\ Rev. A~\textbf{94}, 052508 (2016).
\bibitem{Feynman49} R.P.~Feynman,
   The theory of positrons.
   Phys.\ Rev.\ \textbf{76}, 749 (1949).
\bibitem{Feynman53} M.~Baranger, H.A.~Bethe, R.P.~Feynman,
   Relativistic Correction to the Lamb Shift.
   Phys.\ Rev.\ \textbf{92}, 482 (1953).
\bibitem{Pachucki04} K.~Pachucki,
   Long-wavelength quantum electrodynamics.
   Phys. Rev. A~\textbf{69}, 052502 (2004).
\bibitem{Pachucki05} K.~Pachucki,
   Higher-order effective Hamiltonian for light atomic systems.
   Phys. Rev. A~\textbf{71}, 012503 (2005).
\bibitem{Pachucki98} K.~Pachucki,
   Simple derivation of helium Lamb shift.
   J. Phys. B: At.\ Mol.\ Opt.\ Phys.~\textbf{31}, 5123 (1998).
\bibitem{Pachucki97} K.~Pachucki,
   Effective Hamiltonian approach to the bound state: Positronium hyperfine structure.
   Phys. Rev. A~\textbf{56}, 297 (1997).
\bibitem{Korobov07} V.I.~Korobov and Ts.~Tsogbayar,
   Relativistic corrections of order $m\alpha^6$ to the two-centre problem.
   J.~Phys.~B: At.\ Mol.\ Opt.\ Phys.~\textbf{40}, 2661 (2007).
\bibitem{Korobov08} V.I.~Korobov,
   Relativistic corrections of $m\alpha^6$ order to the rovibrational spectrum of H$_2^+$ and HD$^+$ molecular ions.
   Phys. Rev. A~\textbf{77}, 022509 (2008).
\bibitem{Korobov17} V.I.~Korobov, L.~Hilico, and J.-Ph.~Karr,
   Fundamental Transitions and Ionization Energies of the Hydrogen Molecular Ions with Few ppt Uncertainty.
   Phys. Rev. Lett.~\textbf{118}, 233001 (2017).
\bibitem{Zhong18} Z.-X.~Zhong, W.-P.~Zhou, and X.-S.~Mei,
   Spin-averaged effective Hamiltonian of orders $m\alpha^6$ and $m\alpha^6(m/M)$ for hydrogen molecular ions.
   Phys. Rev. A~\textbf{98}, 032502 (2018).
\bibitem{SapYen} J.R.~Sapirstein, D.R.~Yennie, in: T.~Kinoshita (Ed.),
   {\em Quantum Electrodynamics},
   World Scientific, Singapore, 1990.
\bibitem{EGS} M.I.~Eides, H.~Grotch, and V.A.~Shelyuto,
   {\em Theory of Light Hydrogenic Bound States}.
   Springer Tracts in Modern Physics \textbf{222} (Springer, Berlin, 2007).
\bibitem{Zemach} A.C.~Zemach,
   Proton structure and the hyperfine shift in hydrogen.
   Phys.\ Rev.\ \textbf{104}, 1771 (1956).
\bibitem{Arnowitt} R.~Arnowitt,
   The Hyperfine Structure of Hydrogen.
   Phys.\ Rev.\ \textbf{92}, 1002 (1953).
\bibitem{BodYen88} G.T.~Bodwin and D.R.~Yennie,
   Some corrections to the hydrogen hyperfine splitting.
   Phys.\ Rev.~D \textbf{37}, 498 (1988).
\bibitem{Korobov09b} V.I.~Korobov and Z.-X.~Zhong,
   Spin-orbit corrections of order $m\alpha^6$ to the fine structure of the (37,35) state in the $^4$He$\,\bar{p}$ antiprotonic helium atom.
   Phys. Rev. A~\textbf{80}, 042506 (2009).
\bibitem{Zhong15} Z.-X.~Zhong, X.~Tong, Z.-C.~Yan, and T.-Y.~Shi,
   High-precision spectroscopy of hydrogen molecular ions.
   Chin. Phys. B~\textbf{24}, 053102 (2015).
\bibitem{Korobov00} V.I.~Korobov,
   Coulomb three-body bound-state problem: Variational calculations of nonrelativistic energies.
   Phys. Rev. A~\textbf{61}, 064503 (2000).
\bibitem{Drake06} G.W.F.~Drake,
   High Precision Calculations for Helium,
   in {\em Springer Handbook of Atomic, Molecular and Optical Physics}, ed. G.W.F. Drake (Springer, New York, 2006).
\bibitem{Drake78} G.W.F.~Drake,
   Angular integrals and radial recurrence relations for two-electron matrix elements in Hylleraas coordinates.
   Phys. Rev. A~\textbf{18}, 820 (1978).
\bibitem{Harris04} F.E.~Harris, A.M.~Frolov, and V.H.~Smith, Jr.,
   Singular and nonsingular three-body integrals for exponential wave functions.
   J. Chem. Phys.~\textbf{121}, 6323 (2004).
\bibitem{Pachucki99} K.~Pachucki,
   Quantum electrodynamics effects on helium fine structure.
   J. Phys. B: At.\ Mol.\ Opt.\ Phys.~\textbf{32}, 137 (1999).
\bibitem{Jentschura05} U.D.~Jentschura, A.~Czarnecki, and K.~Pachucki,
   Nonrelativistic QED approach to the Lamb shift.
   Phys. Rev. A~\textbf{72}, 062102 (2005).
\bibitem{Hori16} M.~Hori {\textit et al.},
   Buffer-gas cooling of antiprotonic helium to 1.5 to 1.7 K, and antiproton-to-electron mass ratio.
   Science~\textbf{354}, 610 (2016).
\bibitem{Varsh} D.A.~Varshalovich,  A.N.~Moskalev, and  V.K.~Khersonskii,
  \emph{Quantum Theory of Angular Momentum},
  (Nauka, Leningrad, 1975; World Scientific, Singapore, 1988).

\end{thebibliography}
\end{document}